\documentclass[10pt,nofootinbib]{revtex4-2}
\usepackage[utf8]{inputenc}
\usepackage{color}
\usepackage{calc}
\usepackage{mathrsfs}
\usepackage{mathtools}
\usepackage{amsmath}
\usepackage{amssymb}
\usepackage{graphicx}
\usepackage{microtype}
\usepackage[bookmarks=false,
 breaklinks=false,pdfborder={0 0 1},backref=section,colorlinks=true]
 {hyperref}
\hypersetup{
 linkcolor=blue,citecolor=blue,urlcolor=blue}

\makeatletter
\usepackage{bm}
\usepackage{mathrsfs}
\usepackage{hyphenat}
\usepackage{xcolor}

\allowdisplaybreaks
\numberwithin{equation}{section}

\newcommand{\dd}{\mathrm{d}}
\newcommand{\order}{\mathcal{O}}

\makeatother

\begin{document}
\title{Conserved Gravitational Charges as Curvature Fluxes}
\author{Emel Altas}
\email{emel.altas@agu.edu.tr}

\affiliation{ Department of Engineering Sciences, Abdullah Gül University, 38080
Kayseri, Türkiye }
\author{Bayram Tekin}
\email{bayram.tekin@bilkent.edu.tr}

\affiliation{ Department of Physics, Bilkent University, 06800 Ankara, Türkiye }
\date{Draft: \today}
\begin{abstract}
Conserved gravitational charges are commonly expressed as surface
integrals of the metric perturbation and its first derivatives. We
show that, in the usual asymptotically AdS and asymptotically flat
settings, the standard charges admit equivalent representatives as
fluxes of linearized curvature. The construction uses a divergence-free
rank-four tensor whose trace is proportional to the cosmological Einstein
tensor. On a maximally symmetric AdS background, it reproduces the
known curvature representation of the Abbott--Deser charges. The
asymptotically flat construction is not obtained by taking the naive
$\Lambda\to0$ limit, since the Killing two-form vanishes for translations.
We instead introduce an antisymmetric Poincaré Killing potential.
A representative Killing potential adapted to the algebraic Bianchi
identity converts the linearized Einstein current into a total divergence
and yields a single curvature-flux formula. Its translation sector
reproduces the ADM energy--momentum, while in four dimensions its
Lorentz sector reproduces the angular momentum and boost/center-of-mass
charges under the standard Regge--Teitelboim falloff and parity conditions.
The normalization is checked explicitly for Schwarzschild, boosted
Schwarzschild, and Kerr data. On non-maximally symmetric Einstein
backgrounds, background Weyl curvature generates additional terms,
so the maximally symmetric construction does not directly extend to
a pure codimension-two curvature-flux formula. 
\end{abstract}
\maketitle

\section{Introduction}

Conserved gravitational charges are most commonly written as surface
integrals of the metric perturbation and its first derivatives. In
asymptotically flat gravity, this is the familiar Arnowitt-Deser-Misner
(ADM) construction \cite{ADM}, while on asymptotically anti-de Sitter
backgrounds the corresponding expression is the Abbott--Deser (AD)
charge \cite{AbbottDeser}. These charges are geometric and gauge
invariant once the asymptotic phase space and the allowed transformations
have been fixed, although their standard local representatives are
written in terms of the metric deviation $h_{\mu\nu}$ and $\bar{\nabla}_{\rho}h_{\mu\nu}$.
The question addressed here is whether the same canonical charges
can be represented directly as fluxes of the linearized curvature.

Curvature-based expressions for asymptotic gravitational charges have
a substantial history. The treatment of spatial infinity by Ashtekar
and Hansen provides a covariant curvature-based description of the
asymptotic charges \cite{AshtekarHansen}. In the canonical setting,
Baskaran, Lau, and Petrov studied two-surface integrals linear in
the spacetime Riemann tensor and showed how suitable curvature moments
reproduce the ADM energy and the Beig--Ó Murchadha center of mass
\cite{Petrov1,BeigOMurchadha}. The purpose of the present paper is
therefore not to claim that curvature moments themselves are new.
Our aim is to derive the translation and Lorentz charges from a single
divergence identity based on a rank-four divergence-free tensor and
an antisymmetric potential associated with a Poincaré Killing vector.
The resulting formula gives curvature representatives of the ADM energy,
linear momentum, angular momentum, and boost or center-of-mass charges
within one construction.

The starting point is a tensor with the algebraic symmetries of the
Riemann tensor, whose trace is proportional to the Einstein tensor.
For vanishing cosmological constant, define 
\begin{equation}
\begin{aligned}P^{\mu\nu\rho\sigma}:={}R^{\mu\nu\rho\sigma}+g^{\mu\sigma}R^{\nu\rho}-g^{\mu\rho}R^{\nu\sigma}-g^{\nu\sigma}R^{\mu\rho}+g^{\nu\rho}R^{\mu\sigma}+\frac{R}{2}\left(g^{\mu\rho}g^{\nu\sigma}-g^{\mu\sigma}g^{\nu\rho}\right),\end{aligned}
\label{eq:intro-P-tensor}
\end{equation}
which satisfies 
\begin{equation}
P^{\mu}{}_{\nu\mu\sigma}=-(n-3)G_{\nu\sigma},\qquad G_{\nu\sigma}:=R_{\nu\sigma}-\frac{1}{2}Rg_{\nu\sigma},\label{eq:P-trace-einstein-lowered}
\end{equation}
and is identically divergence-free, 
\begin{equation}
\nabla_{\mu}P^{\mu\nu\rho\sigma}=0.\label{eq:intro-P-divergence-free}
\end{equation}
These properties hold off-shell. The construction below assumes $n>3$,
since the trace relation degenerates in three spacetime dimensions;
indeed, in $n=3$ one has $G_{\mu\nu}=({}^{\star}R^{\star})_{\mu\nu}$,
which is another way to see why the construction fails there. In four
dimensions, $P_{\mu\nu\rho\sigma}$ is the double Hodge dual of the
Riemann tensor 
\begin{equation}
P_{\mu\nu\rho\sigma}=({}^{\star}R^{\star})_{\mu\nu\rho\sigma}.\label{eq:intro-double-dual}
\end{equation}
In dimensions $n>4$, the ordinary Hodge dual maps a two-form to an
$(n-2)$-form, so the direct four-dimensional double-dual statement
no longer applies. The corresponding generalized double-dual expression
is 
\begin{equation}
P^{\mu\nu}{}_{\rho\sigma}=-\frac{1}{4(n-4)!}\,\epsilon^{\mu\nu\alpha\beta\lambda_{1}\cdots\lambda_{n-4}}\,\epsilon_{\rho\sigma\gamma\delta\lambda_{1}\cdots\lambda_{n-4}}\,R^{\gamma\delta}{}_{\alpha\beta}.\label{eq:P-generalized-double-dual}
\end{equation}
For $n=4$, this reduces to \eqref{eq:intro-double-dual}. Equivalently,
for all $n\geq4$, one may write 
\begin{equation}
P^{\mu\nu\rho\sigma}=\frac{1}{2}\frac{\partial}{\partial R_{\mu\nu\rho\sigma}}\left(R_{\alpha\beta\gamma\delta}R^{\alpha\beta\gamma\delta}-4R_{\alpha\beta}R^{\alpha\beta}+R^{2}\right),\label{eq:P-GB-functional-derivative}
\end{equation}
where the metric is kept fixed in the differentiation, and the derivative
is taken with the algebraic symmetries of the Riemann tensor imposed.
Thus $P^{\mu\nu\rho\sigma}$ is one half of the derivative of the
Gauss--Bonnet scalar with respect to the Riemann tensor. This form
is useful for generalizations: differentiating any Lanczos--Lovelock
invariant with respect to the Riemann tensor gives the corresponding
higher-curvature analogue; see, for example, equation (21) of \cite{Ozkarsligil}.

For cosmological Einstein gravity, it is useful to introduce a shifted
tensor whose trace is proportional to the cosmological Einstein tensor.
We denote it by $\mathcal{P}^{\nu}{}_{\mu\beta\sigma}$, reserving
$P^{\mu\nu\rho\sigma}$ for the tensor in \eqref{eq:intro-P-tensor}:
\begin{equation}
\begin{aligned}\mathcal{P}^{\nu}{}_{\mu\beta\sigma}:=R^{\nu}{}_{\mu\beta\sigma}+\delta_{\sigma}^{\nu}\mathcal{G}_{\beta\mu}-\delta_{\beta}^{\nu}\mathcal{G}_{\sigma\mu}+\mathcal{G}^{\nu}{}_{\sigma}g_{\beta\mu}-\mathcal{G}^{\nu}{}_{\beta}g_{\sigma\mu}+\left(\frac{R}{2}-\frac{\Lambda(n+1)}{n-1}\right)\left(\delta_{\sigma}^{\nu}g_{\beta\mu}-\delta_{\beta}^{\nu}g_{\sigma\mu}\right),\end{aligned}
\label{eq:P-cosmological-definition}
\end{equation}
where 
\begin{equation}
\mathcal{G}_{\mu\nu}:=R_{\mu\nu}-\frac{1}{2}Rg_{\mu\nu}+\Lambda g_{\mu\nu}.\label{eq:cosmo-Einstein-tensor}
\end{equation}
This tensor obeys 
\begin{equation}
\mathcal{P}^{\nu}{}_{\mu\nu\sigma}=-(n-3)\mathcal{G}_{\mu\sigma},\label{eq:P-cosmological-trace}
\end{equation}
together with the off-shell identity 
\begin{equation}
\nabla_{\nu}\mathcal{P}^{\nu\mu\beta\sigma}=0.\label{eq:intro-cosmological-P-divergence}
\end{equation}
The constant term in \eqref{eq:P-cosmological-definition} is chosen
so that 
\begin{equation}
\bar{\mathcal{P}}^{\nu}{}_{\mu\beta\sigma}=0\label{eq:P-vanishes-AdS}
\end{equation}
on a maximally symmetric background satisfying 
\begin{equation}
\bar{R}_{\mu\alpha\nu\beta}=\frac{2\Lambda}{(n-1)(n-2)}\left(\bar{g}_{\mu\nu}\bar{g}_{\alpha\beta}-\bar{g}_{\mu\beta}\bar{g}_{\alpha\nu}\right).\label{eq:maxsym-Riemann}
\end{equation}
The vanishing of $\bar{\mathcal{P}}^{\nu}{}_{\mu\beta\sigma}$ is
the central simplification in the AdS construction. It makes $(\mathcal{P}^{\nu}{}_{\mu\beta\sigma})^{(1)}$
invariant under linearized background diffeomorphisms. The second
ingredient is the Killing two-form 
\begin{equation}
\bar{S}_{\beta\sigma}:=\bar{\nabla}_{\beta}\bar{\xi}_{\sigma}=-\bar{\nabla}_{\sigma}\bar{\xi}_{\beta}.\label{eq:intro-Killing-two-form}
\end{equation}
On a maximally symmetric background, the Killing identity relates
$\bar{\nabla}_{\nu}\bar{S}_{\beta\sigma}$ algebraically to $\bar{\xi}^{\mu}$.
Combining this fact with the trace and divergence identities of $\mathcal{P}^{\nu}{}_{\mu\beta\sigma}$
gives the curvature representation of the Abbott--Deser charge obtained
in ~\cite{Altas1,Altas2}. In particular, the charge can be evaluated
at infinity from the linearized Riemann tensor rather than from a
metric-potential representative.

The asymptotically flat construction requires a different step. It
is not the direct $\Lambda\to0$ limit of the AdS formula. For a translation
$\xi^{\mu}=a^{\mu}$ in Minkowski space, 
\begin{equation}
\partial_{\mu}\xi_{\nu}=0,\label{eq:translation-killing-two-form-zero}
\end{equation}
and hence the Killing two-form used in AdS vanishes. We replace it
by an antisymmetric Killing potential satisfying 
\begin{equation}
F^{\mu\nu}[\xi]=-F^{\nu\mu}[\xi],\qquad\partial_{\nu}F^{\nu\mu}[\xi]=\xi^{\mu}.\label{eq:intro-flat-killing-potential}
\end{equation}
For translations, this potential is affine in the Cartesian coordinates.
For a general Poincaré Killing vector, it may be chosen in a form adapted
to the algebraic Bianchi identity of the $P$-tensor. Inserting this
representative into the exact divergence identity converts the linearized
Einstein current into a total divergence and produces one curvature-flux
formula for the full Poincaré family of generators.

The translation sector of this formula gives the ADM energy and linear
momentum. We reduce the corresponding curvature expressions explicitly
to the standard ADM surface integrals and verify the normalization
for Schwarzschild and boosted Schwarzschild data. In four dimensions,
the Lorentz sector gives the Regge--Teitelboim angular momentum and
boost generators under the usual falloff and parity conditions \cite{ReggeTeitelboim}.
The boost charge yields the center of mass when the ADM energy is
nonzero, while the rotational formula reproduces the spin of Kerr.
The additional term in the Lorentz Killing potential is essential:
it is precisely the term that allows the Lorentz-dependent part of
the $P$-tensor contraction to vanish by the algebraic Bianchi identity.

It is also useful to identify the limitations of the AdS argument.
On an Einstein background that is not maximally symmetric, $\bar{\mathcal{P}}^{\nu}{}_{\mu\beta\sigma}$
does not vanish; instead, its background value is the Weyl tensor,
\begin{equation}
\bar{\mathcal{P}}_{\mu\nu\rho\sigma}=\bar{C}_{\mu\nu\rho\sigma}.\label{eq:intro-Pbar-Weyl}
\end{equation}
This identity refers to $\mathcal{P}$ as defined in \eqref{eq:P-cosmological-definition};
on Einstein backgrounds, it holds in any dimension $n>3$. It should
not be read as a statement about double duals: for $n>4$ the generalized
double dual of the Weyl tensor is not the Weyl tensor. Linearizing
the exact divergence identity then generates terms involving the background
Weyl curvature, including the contributions arising from the nonzero
background value of $\mathcal{P}$. These terms vanish in the maximally
symmetric case but prevent a direct repetition of the AdS derivation
on a generic Einstein background. We use this observation only as
a diagnostic statement; no general local, gauge-invariant, curvature-only
superpotential for the background-Weyl contributions is claimed here.

The paper is organized as follows: Section~2 fixes the curvature,
normalization, and orientation conventions. Section~3 reviews the
Killing current and the Abbott--Deser metric-potential representative.
Section~4 recalls the AdS curvature-flux construction. Sections~5
and~6 introduce the flat-space Killing potential and derive the master
curvature-flux identity. Section~7 treats the ADM energy and linear
momentum and gives the Schwarzschild and boosted-Schwarzschild checks.
Sections~8 and~9 analyze the Lorentz sector, including rotations,
boosts, the center of mass, the asymptotic Poincaré algebra, and the
Kerr example. Section~10 discusses gauge invariance and boundary
conditions. Section~11 explains the additional terms that arise on
non-maximally symmetric Einstein backgrounds. Section~12 contains
the conclusions, and Appendix~A comments on the canonical interpretation
of the AdS charge algebra. 

\section{Conventions}

\label{sec:conventions} We use the conventions of \cite{AdamiReview}.
The metric signature is mostly plus, 
\begin{equation}
g_{\mu\nu}=\mathrm{diag}(-,+,+,\ldots,+),\label{eq:signature}
\end{equation}
locally in an orthonormal frame. Greek indices run over the spacetime
values $0,1,\ldots,n-1$, while Latin indices run over the spatial
values $1,\ldots,n-1$. The Riemann tensor is defined by 
\begin{equation}
[\nabla_{\mu},\nabla_{\nu}]V^{\rho}=R^{\rho}{}_{\sigma\mu\nu}V^{\sigma}.\label{eq:Riemann-convention}
\end{equation}
The Ricci tensor and the scalar curvature are 
\begin{equation}
R_{\mu\nu}=R^{\rho}{}_{\mu\rho\nu},\qquad R=g^{\mu\nu}R_{\mu\nu}.\label{eq:Ricci-scalar}
\end{equation}
The cosmological Einstein tensor is denoted by 
\begin{equation}
\mathcal{G}_{\mu\nu}:=R_{\mu\nu}-\frac{1}{2}Rg_{\mu\nu}+\Lambda g_{\mu\nu}.\label{eq:conventions-cosmological-Einstein-tensor}
\end{equation}
For $\Lambda=0$, this reduces to the ordinary Einstein tensor 
\begin{equation}
G_{\mu\nu}:=R_{\mu\nu}-\frac{1}{2}Rg_{\mu\nu}.\label{eq:ordinary-Einstein-tensor}
\end{equation}
Einstein's equations are normalized as 
\begin{equation}
\mathcal{G}_{\mu\nu}=\kappa_{n}T_{\mu\nu}.\label{eq:field-equation-normalization}
\end{equation}
For $\Lambda=0$, this becomes $G_{\mu\nu}=\kappa_{n}T_{\mu\nu}$.
In four dimensions, 
\begin{equation}
\kappa_{4}=8\pi G.\label{eq:kappa-four-convention}
\end{equation}
More generally, we keep $\kappa_{n}$ explicit. When comparing with
the normalization often used in the $n$-dimensional Killing-charge
literature, one may set 
\begin{equation}
\kappa_{n}=2\Omega_{n-2}G_{n},\label{eq:kappa-n-convention}
\end{equation}
where $\Omega_{n-2}$ is the area of the unit $(n-2)$-sphere.

Background quantities are denoted with a bar. Thus 
\begin{equation}
g_{\mu\nu}=\bar{g}_{\mu\nu}+h_{\mu\nu},\label{eq:background-split}
\end{equation}
and all indices on linearized quantities are raised and lowered with
$\bar{g}_{\mu\nu}$, unless otherwise stated. All linearized curvatures
and linearized field equations are evaluated about the background
metric $\bar{g}_{\mu\nu}$. When an expression such as $(R^{\nu\mu}{}_{\beta\sigma})^{(1)}$
appears, the indices are raised after linearization with the background
metric, unless explicitly stated otherwise.

Antisymmetrization is defined with weight one: 
\begin{equation}
A_{[\mu\nu]}:=\frac{1}{2}\left(A_{\mu\nu}-A_{\nu\mu}\right).\label{eq:antisymmetrization-convention}
\end{equation}

About the flat background, the linearized Riemann tensor used in Secs.~VI--IX
is 
\begin{equation}
\left(R_{\alpha\beta\mu\nu}\right)^{(1)}=\frac{1}{2}\left(\partial_{\mu}\partial_{\beta}h_{\alpha\nu}+\partial_{\nu}\partial_{\alpha}h_{\beta\mu}-\partial_{\nu}\partial_{\beta}h_{\alpha\mu}-\partial_{\mu}\partial_{\alpha}h_{\beta\nu}\right).\label{eq:linearized-riemann-conventions}
\end{equation}
Since the whole construction is sensitive to orientation-dependent
signs, we fix the global normalization once and for all on a reference
solution: with the conventions of this section and the binormal orientation
adopted below, the charge of the Schwarzschild solution is $E=+M$.

\section{Killing charges in metric-perturbation form}

\label{sec:metric-perturbation-charges}

Let the metric field equations of a diffeomorphism-invariant gravity
theory be written as 
\begin{equation}
E_{\mu\nu}[g]=\kappa_{n}\tau_{\mu\nu},\qquad\nabla_{\mu}E^{\mu\nu}[g]=0.\label{eq:generic-field-equation}
\end{equation}
Here $E_{\mu\nu}[g]$ denotes the geometric Euler--Lagrange tensor,
while $\tau_{\mu\nu}$ denotes the matter stress tensor, when matter
is present. The second relation is the off-shell generalized Bianchi
identity following from diffeomorphism invariance of the gravitational
action.

Let $\bar{g}_{\mu\nu}$ be an exact vacuum solution, 
\begin{equation}
E_{\mu\nu}[\bar{g}]=0,\label{eq:background-equation}
\end{equation}
and let $\bar{\xi}^{\mu}$ be a Killing vector of the background,
\begin{equation}
\bar{\nabla}_{(\mu}\bar{\xi}_{\nu)}=0.\label{eq:background-Killing}
\end{equation}
For (\ref{eq:background-split}) define the nonlinear remainder 
\begin{equation}
\Delta E_{\mu\nu}[h]:=E_{\mu\nu}[\bar{g}+h]-E_{\mu\nu}[\bar{g}]-\left(E_{\mu\nu}\right)^{(1)}.\label{eq:nonlinear-field-equation-remainder}
\end{equation}
The exact field equations may then be written as 
\begin{equation}
\left(E_{\mu\nu}\right)^{(1)}=\kappa_{n}T_{\mu\nu}^{\mathrm{eff}},\label{eq:linearized-equation}
\end{equation}
where 
\begin{equation}
T_{\mu\nu}^{\mathrm{eff}}:=\tau_{\mu\nu}-\frac{1}{\kappa_{n}}\Delta E_{\mu\nu}[h].\label{eq:effective-source-definition}
\end{equation}
Thus $T_{\mu\nu}^{\mathrm{eff}}$ contains the matter stress tensor
together with the terms that are nonlinear in the metric perturbation,
see \cite{DeserTekinPRL,DeserTekinPRD}. Linearizing the generalized
Bianchi identity about a background satisfying \eqref{eq:background-equation}
gives 
\begin{equation}
\bar{\nabla}_{\mu}\left(E^{\mu\nu}\right)^{(1)}=0.\label{eq:linearized-Bianchi}
\end{equation}
Every background Killing vector, therefore, defines the conserved
current 
\begin{equation}
J^{\mu}[\bar{\xi}]:=\bar{\xi}_{\nu}\left(E^{\mu\nu}\right)^{(1)}.\label{eq:linear-current-definition}
\end{equation}
Indeed, one has 
\begin{align}
\bar{\nabla}_{\mu}J^{\mu}[\bar{\xi}] & =\left(\bar{\nabla}_{\mu}\bar{\xi}_{\nu}\right)\left(E^{\mu\nu}\right)^{(1)}+\bar{\xi}_{\nu}\bar{\nabla}_{\mu}\left(E^{\mu\nu}\right)^{(1)}=0.\label{eq:linear-current-conservation}
\end{align}
The first term vanishes because $\bar{\nabla}_{\mu}\bar{\xi}_{\nu}$
is antisymmetric and $(E^{\mu\nu})^{(1)}$ is symmetric, while the
second vanishes by \eqref{eq:linearized-Bianchi}.

Let $\bar{\Sigma}$ be a spacelike hypersurface in the background
spacetime. The linearized charge associated with $\bar{\xi}^{\mu}$
is 
\begin{equation}
Q[\bar{\xi};\bar{\Sigma}]:=\frac{1}{\kappa_{n}}\int_{\bar{\Sigma}}\dd\bar{\Sigma}_{\mu}\,J^{\mu}[\bar{\xi}]=\frac{1}{\kappa_{n}}\int_{\bar{\Sigma}}\dd\bar{\Sigma}_{\mu}\,\bar{\xi}_{\nu}\left(E^{\mu\nu}\right)^{(1)}.\label{eq:general-charge-volume}
\end{equation}
Current conservation implies that this charge is independent of the
choice of hypersurface, provided that the flux of $J^{\mu}[\bar{\xi}]$
through the intervening timelike boundary vanishes. This requirement
is part of the boundary conditions defining the asymptotic phase space.

To express \eqref{eq:general-charge-volume} as a codimension-two
surface integral, one seeks an antisymmetric tensor $\mathcal{F}^{\mu\nu}=-\mathcal{F}^{\nu\mu}$
satisfying 
\begin{equation}
J^{\mu}[\bar{\xi}]=\bar{\nabla}_{\nu}\mathcal{F}^{\mu\nu}.\label{eq:current-superpotential-general}
\end{equation}
Such a representative is not unique. One may add an identically conserved
improvement without changing the integrated charge under the assumed
boundary conditions.

We now specialize in cosmological Einstein gravity, 
\begin{equation}
E_{\mu\nu}=\mathcal{G}_{\mu\nu}:=R_{\mu\nu}-\frac{1}{2}Rg_{\mu\nu}+\Lambda g_{\mu\nu},\label{eq:section-cosmological-einstein-tensor}
\end{equation}
and take the background to satisfy 
\begin{equation}
\bar{\mathcal{G}}_{\mu\nu}=0.\label{eq:Einstein-background-equation}
\end{equation}
The background is therefore an Einstein spacetime, 
\begin{equation}
\bar{R}_{\mu\nu}=\frac{2\Lambda}{n-2}\,\bar{g}_{\mu\nu},\label{eq:Einstein-background-Ricci}
\end{equation}
but it need not yet be maximally symmetric.

The Abbott--Deser identity reads 
\begin{equation}
\bar{\xi}_{\nu}\left(\mathcal{G}^{\mu\nu}\right)^{(1)}=\bar{\nabla}_{\alpha}\mathcal{F}_{\mathrm{AD}}^{\mu\alpha}[h;\bar{\xi}],\qquad\mathcal{F}_{\mathrm{AD}}^{\mu\alpha}=-\mathcal{F}_{\mathrm{AD}}^{\alpha\mu}.\label{eq:AD-divergence}
\end{equation}
To write the potential compactly, introduce the shifted perturbation
(trace-reversed perturbation in four dimensions) 
\begin{equation}
\widetilde{h}^{\mu\nu}:=h^{\mu\nu}-\frac{1}{2}\bar{g}^{\mu\nu}h,\qquad h:=\bar{g}^{\rho\sigma}h_{\rho\sigma},\label{eq:h-trace-reversed}
\end{equation}
and the rank-four tensor 
\begin{equation}
\mathscr{K}^{\mu\alpha\nu\beta}:=\frac{1}{2}\left(\bar{g}^{\alpha\nu}\widetilde{h}^{\mu\beta}+\bar{g}^{\mu\beta}\widetilde{h}^{\alpha\nu}-\bar{g}^{\alpha\beta}\widetilde{h}^{\mu\nu}-\bar{g}^{\mu\nu}\widetilde{h}^{\alpha\beta}\right).\label{eq:K-superpotential}
\end{equation}
It has the algebraic symmetries 
\begin{equation}
\mathscr{K}^{\mu\alpha\nu\beta}=-\mathscr{K}^{\alpha\mu\nu\beta}=-\mathscr{K}^{\mu\alpha\beta\nu}=\mathscr{K}^{\nu\beta\mu\alpha}.\label{eq:K-superpotential-symmetries}
\end{equation}
The Abbott--Deser two-form is then 
\begin{equation}
\mathcal{F}_{\mathrm{AD}}^{\mu\alpha}[h;\bar{\xi}]=\bar{\xi}_{\nu}\bar{\nabla}_{\beta}\mathscr{K}^{\mu\alpha\nu\beta}-\mathscr{K}^{\mu\beta\nu\alpha}\bar{\nabla}_{\beta}\bar{\xi}_{\nu}.\label{eq:AD-two-form}
\end{equation}
Its antisymmetry follows from \eqref{eq:K-superpotential-symmetries}
and the Killing equation. Using the Stokes' theorem, the corresponding
surface charge is 
\begin{equation}
Q_{\mathrm{AD}}[\bar{\xi}]=\frac{1}{\kappa_{n}}\int_{\partial\bar{\Sigma}}\dd\bar{\Sigma}_{\mu\alpha}\,\mathcal{F}_{\mathrm{AD}}^{\mu\alpha}[h;\bar{\xi}].\label{eq:AD-charge}
\end{equation}

\begin{figure}[h!]
\centering \includegraphics[width=0.5\columnwidth]{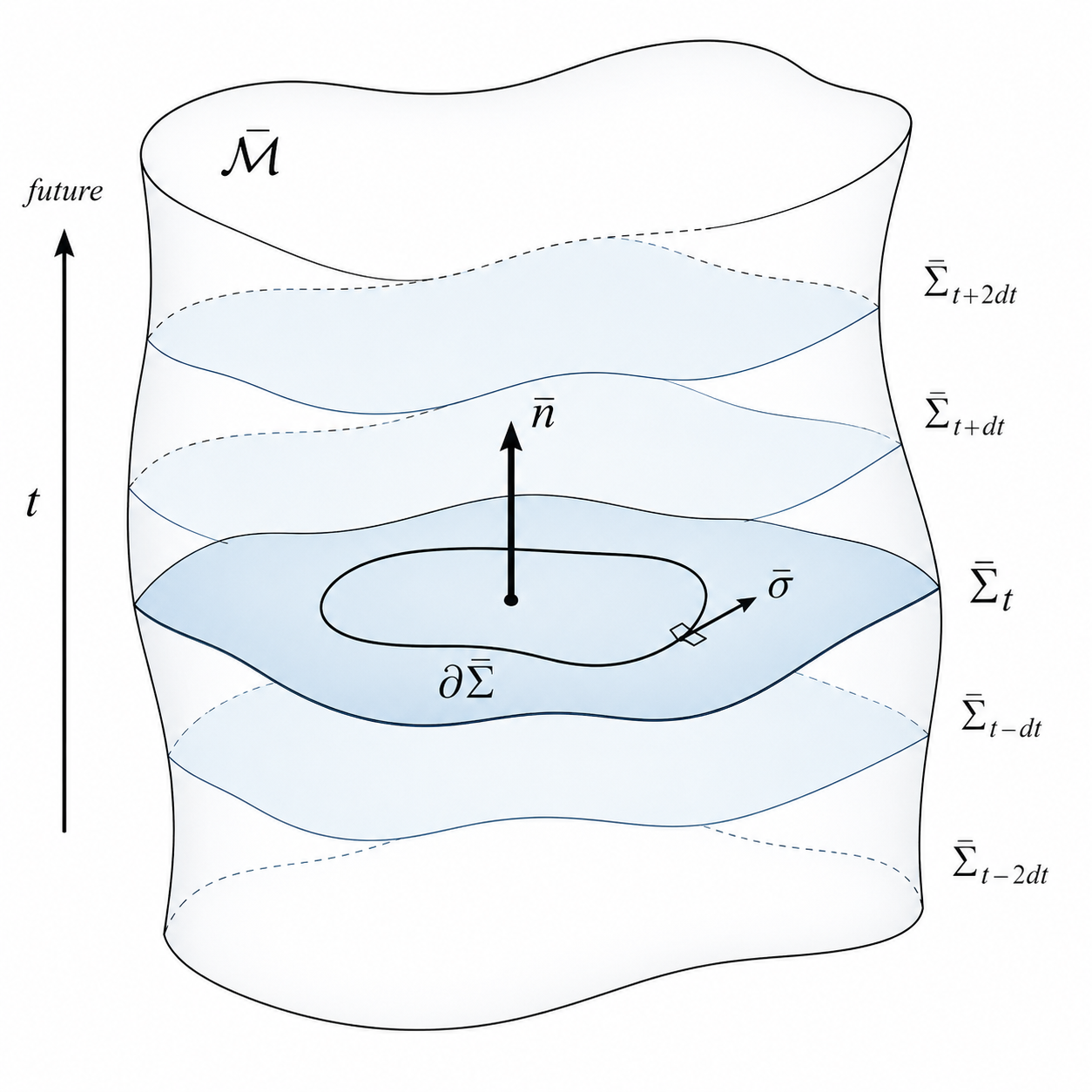}
\caption{ Geometric setup for the charge integral. The spacelike hypersurface
$\bar{\Sigma}$ has future-directed unit normal $\bar{n}^{\mu}$,
and its codimension-two boundary $\partial\bar{\Sigma}$ has outward-pointing
unit normal $\bar{\sigma}^{\mu}$ within $\bar{\Sigma}$. The orientation
is as $\dd\bar{\Sigma}_{\mu\nu}=\bar{n}_{[\mu}\bar{\sigma}_{\nu]}\dd\bar{S}$. }
\label{fig:foliation-geometry} 
\end{figure}

Equation~\eqref{eq:AD-charge} is the standard metric-potential representative
of the conserved charge. It is written in terms of $h_{\mu\nu}$ and
its first background derivative. Under the admissible asymptotic gauge
transformations, its variation is an improvement term and the integrated
charge is unchanged.

It is important to separate two statements. The Abbott--Deser identity
\eqref{eq:AD-divergence} is valid on an Einstein background admitting
a Killing vector; maximal symmetry is not required at this stage.
Maximal symmetry becomes essential in the next section, where we seek
an equivalent representative written directly as a flux of linearized
curvature. On a maximally symmetric AdS background, the divergence-free
$\mathcal{P}$-tensor provides precisely the bridge between the Abbott--Deser
current and such a curvature-flux representative. 

\section{Charges as curvature fluxes in AdS}

\label{sec:AdS-curvature-flux}

We first recall the AdS curvature-flux construction of \cite{Altas1,Altas2}.
Besides fixing our notation, this discussion shows why the corresponding
asymptotically flat formula cannot be obtained by taking a naive $\Lambda\to0$
limit.

Let $\bar{g}_{\mu\nu}$ be an $n$-dimensional AdS background, 
\begin{equation}
\bar{R}_{\mu\nu\rho\sigma}=\frac{2\Lambda}{(n-1)(n-2)}\left(\bar{g}_{\mu\rho}\bar{g}_{\nu\sigma}-\bar{g}_{\mu\sigma}\bar{g}_{\nu\rho}\right),\qquad\Lambda<0,\label{eq:AdS-constant-curvature}
\end{equation}
and let $\bar{\xi}^{\mu}$ be a Killing vector, 
\begin{equation}
\bar{\nabla}_{(\mu}\bar{\xi}_{\nu)}=0.\label{eq:AdS-background-Killing}
\end{equation}
We assume $n>3$, since the trace of the $\mathcal{P}$-tensor degenerates
in three spacetime dimensions.

Define the antisymmetric Killing two-form 
\begin{equation}
\bar{S}_{\beta\sigma}:=\bar{\nabla}_{\beta}\bar{\xi}_{\sigma}=-\bar{\nabla}_{\sigma}\bar{\xi}_{\beta}.\label{eq:AdS-Killing-two-form}
\end{equation}
The starting point is the exact identity 
\begin{equation}
\nabla_{\nu}\left(F^{\beta\sigma}\mathcal{P}^{\nu\mu}{}_{\beta\sigma}\right)=\mathcal{P}^{\nu\mu}{}_{\beta\sigma}\nabla_{\nu}F^{\beta\sigma},\label{eq:exact-PF-identity}
\end{equation}
which holds for every antisymmetric tensor $F^{\beta\sigma}$. It
follows from metric compatibility and the off-shell identity 
\begin{equation}
\nabla_{\nu}\mathcal{P}^{\nu\mu}{}_{\beta\sigma}=0.\label{eq:P-divergence-free-used-AdS}
\end{equation}
The shifted tensor defined in \eqref{eq:P-cosmological-definition}
vanishes identically on the AdS background: 
\begin{equation}
\bar{\mathcal{P}}^{\nu\mu}{}_{\beta\sigma}=0.\label{eq:P-background-zero-AdS}
\end{equation}
Consequently, the linearization of \eqref{eq:exact-PF-identity} about
AdS contains no terms proportional to a connection variation acting
on the background value of $\mathcal{P}^{\nu\mu}{}_{\beta\sigma}$.
Taking $F^{\beta\sigma}=\bar{S}^{\beta\sigma}$, one obtains 
\begin{equation}
\bar{\nabla}_{\nu}\left[\bar{S}^{\beta\sigma}\left(\mathcal{P}^{\nu\mu}{}_{\beta\sigma}\right)^{(1)}\right]=\left(\mathcal{P}^{\nu\mu}{}_{\beta\sigma}\right)^{(1)}\bar{\nabla}_{\nu}\bar{S}^{\beta\sigma}.\label{eq:linear-PF-identity-AdS}
\end{equation}
With the curvature convention \eqref{eq:Riemann-convention}, the
Killing identity reads 
\begin{equation}
\bar{\nabla}_{\nu}\bar{S}_{\beta\sigma}=\bar{\nabla}_{\nu}\bar{\nabla}_{\beta}\bar{\xi}_{\sigma}=\bar{R}_{\sigma\beta\nu\lambda}\bar{\xi}^{\lambda}.\label{eq:Killing-second-derivative}
\end{equation}
Equivalently, 
\begin{equation}
\bar{\nabla}_{\nu}\bar{S}^{\beta\sigma}=\bar{R}^{\sigma\beta}{}_{\nu\lambda}\bar{\xi}^{\lambda}.\label{eq:Killing-second-derivative-raised}
\end{equation}
Using \eqref{eq:AdS-constant-curvature}, the contraction on the right-hand
side of \eqref{eq:linear-PF-identity-AdS} becomes 
\begin{align}
 & \left(\mathcal{P}^{\nu\mu}{}_{\beta\sigma}\right)^{(1)}\bar{\nabla}_{\nu}\bar{S}^{\beta\sigma}=\frac{2\Lambda}{(n-1)(n-2)}\left(\mathcal{P}^{\nu\mu}{}_{\beta\sigma}\right)^{(1)}\left(\delta_{\nu}^{\sigma}\delta_{\lambda}^{\beta}-\delta_{\lambda}^{\sigma}\delta_{\nu}^{\beta}\right)\bar{\xi}^{\lambda}=\frac{4\Lambda(n-3)}{(n-1)(n-2)}\bar{\xi}_{\lambda}\left(\mathcal{G}^{\mu\lambda}\right)^{(1)}.\label{eq:AdS-P-contraction}
\end{align}
In the last step, we used 
\begin{equation}
\left(\mathcal{P}^{\nu\mu}{}_{\nu\sigma}\right)^{(1)}=-(n-3)\left(\mathcal{G}^{\mu}{}_{\sigma}\right)^{(1)}.\label{eq:linear-P-trace-AdS}
\end{equation}
Equations~\eqref{eq:linear-PF-identity-AdS} and \eqref{eq:AdS-P-contraction}
therefore give 
\begin{equation}
\bar{\xi}_{\lambda}\left(\mathcal{G}^{\mu\lambda}\right)^{(1)}=\frac{(n-1)(n-2)}{4\Lambda(n-3)}\bar{\nabla}_{\nu}\left[\left(\mathcal{P}^{\nu\mu}{}_{\beta\sigma}\right)^{(1)}\bar{\nabla}^{\beta}\bar{\xi}^{\sigma}\right].\label{eq:AdS-current-identity}
\end{equation}
This is the basic AdS curvature identity. It expresses the Abbott--Deser
Killing current as the divergence of a two-form constructed from the
linearized $\mathcal{P}$-tensor. Integrating \eqref{eq:AdS-current-identity}
over a spacelike hypersurface $\bar{\Sigma}$, gives 
\begin{equation}
Q[\bar{\xi}]=\frac{(n-1)(n-2)}{4\kappa_{n}\Lambda(n-3)}\int_{\partial\bar{\Sigma}}\dd\bar{\Sigma}_{\mu\nu}\,\left(\mathcal{P}^{\nu\mu}{}_{\beta\sigma}\right)^{(1)}\bar{\nabla}^{\beta}\bar{\xi}^{\sigma}.\label{eq:AdS-P-charge}
\end{equation}
The order $\nu\mu$ in the integrand is intentional and is tied to
the binormal convention $\dd\bar{\Sigma}_{\mu\nu}=\bar{n}_{[\mu}\bar{\sigma}_{\nu]}\dd\bar{S}$.
Equation~\eqref{eq:AdS-P-charge} is the exact linearized $\mathcal{P}$-tensor
representative; no asymptotic field equation has yet been used.

To obtain the Riemann-flux form, it is useful to retain the mixed
index placement appearing in \eqref{eq:AdS-P-charge}. Linearizing
the definition of the shifted tensor about AdS gives 
\begin{align}
\left(\mathcal{P}^{\nu\mu}{}_{\beta\sigma}\right)^{(1)}={}\left(R^{\nu\mu}{}_{\beta\sigma}\right)^{(1)}+2\delta_{[\sigma}^{\nu}\left(\mathcal{G}_{\beta]}{}^{\mu}\right)^{(1)}+2\delta_{[\beta}^{\mu}\left(\mathcal{G}^{\nu}{}_{\sigma]}\right)^{(1)}+\left(R\right)^{(1)}\delta_{[\sigma}^{\nu}\delta_{\beta]}^{\mu}.\label{eq:P-linearized-AdS-expansion}
\end{align}
This identity is preferable to the corresponding all-up expression:
because the last two indices remain lowered, no separate explicit
$\Lambda h_{\mu\nu}$ term appears.

For an asymptotically AdS solution, the falloff conditions and the
asymptotic vacuum equations imply that the terms involving $(\mathcal{G}_{\mu\nu})^{(1)}$
and $R^{(1)}$ decay too rapidly to contribute to the boundary integral.
Hence the charge may be evaluated from the mixed-index linearized
Riemann tensor: 
\begin{equation}
Q[\bar{\xi}]=\frac{(n-1)(n-2)}{4\kappa_{n}\Lambda(n-3)}\int_{\partial\bar{\Sigma}}\dd\bar{S}\,\bar{n}_{\mu}\bar{\sigma}_{\nu}\left(R^{\nu\mu}{}_{\beta\sigma}\right)^{(1)}\bar{\nabla}^{\beta}\bar{\xi}^{\sigma}.\label{eq:AdS-Riemann-flux-charge-binormal}
\end{equation}

Two points should be kept distinct. Equation~\eqref{eq:AdS-P-charge}
is an exact identity at linear order, whereas \eqref{eq:AdS-Riemann-flux-charge-binormal}
is its asymptotic Riemann representative, obtained after imposing
the AdS falloff conditions. Moreover, the choice of $\bar{\xi}^{\mu}$
selects the energy, angular momentum, or other AdS isometry charge,
but the underlying curvature-flux expression is the same for every
AdS Killing vector. See the applications of the last formula in \cite{Altas4}
and a detailed discussion about the virtues of using the curvature
in the conserved charges in \cite{Ogurol}.

The curvature-flux formula does not define a new conserved quantity.
It is an alternative representative of the standard Abbott--Deser
charge. Under the same asymptotic conditions, 
\begin{equation}
Q[\bar{\xi}]=Q_{{\rm AD}}[\bar{\xi}].\label{eq:AdS-charge-equivalence}
\end{equation}
The difference between the metric-potential and curvature representatives
is an allowed boundary improvement and does not change the integrated
charge.

Finally, both $n>3$ and $\Lambda\neq0$ are essential in \eqref{eq:AdS-current-identity}.
The formula therefore has neither a direct three-dimensional limit
nor a direct asymptotically flat limit. For a flat translation $\xi^{\mu}=a^{\mu}$,
$\partial_{\mu}\xi_{\nu}=0$, so the Killing two-form used above vanishes.
The asymptotically flat construction consequently requires a different
antisymmetric potential, to which we now turn.

\section{Flat-space Killing potentials}

\label{sec:flat-Killing-potentials}

As we have seen above, the AdS construction uses the Killing two-form
\begin{equation}
\bar{S}_{\mu\nu}:=\bar{\nabla}_{\mu}\bar{\xi}_{\nu}=-\bar{\nabla}_{\nu}\bar{\xi}_{\mu}.\label{eq:AdS-Killing-two-form-reminder}
\end{equation}
On a constant-curvature background, $\bar{\nabla}_{\rho}\bar{S}_{\mu\nu}$
is algebraically related to the Killing vector through the background
Riemann tensor. For a translation in flat space, however, 
\begin{equation}
\xi^{\mu}=a^{\mu},\qquad\partial_{\mu}a^{\nu}=0,\label{eq:flat-translation-vector}
\end{equation}
and therefore 
\begin{equation}
\partial_{\mu}\xi_{\nu}=0.\label{eq:flat-translation-gradient-zero}
\end{equation}
The Killing two-form consequently vanishes and cannot encode the translational
ADM charges.

For the flat-space construction, we instead use an antisymmetric Killing
potential, 
\begin{equation}
F^{\mu\nu}[\xi]=-F^{\nu\mu}[\xi],\qquad\partial_{\nu}F^{\nu\mu}[\xi]=\xi^{\mu}.\label{eq:flat-Killing-potential-definition}
\end{equation}
Such potentials are not unique: if 
\begin{equation}
\partial_{\nu}H^{\nu\mu}=0,\qquad H^{\mu\nu}=-H^{\nu\mu},\label{eq:flat-Killing-potential-improvement}
\end{equation}
then $F^{\mu\nu}+H^{\mu\nu}$ satisfies the same defining equation.
The term ``Killing potential'' is used here in the sense of \cite{KastorRayTraschen}.
In Cartesian Minkowski coordinates, a general Poincaré Killing vector
is 
\begin{equation}
\xi^{\mu}=a^{\mu}+\omega^{\mu}{}_{\nu}x^{\nu},\qquad\omega_{\mu\nu}=-\omega_{\nu\mu},\label{eq:flat-Poincare-Killing}
\end{equation}
where $a^{\mu}$ generates translations and $\omega_{\mu\nu}$ generates
Lorentz transformations. Its derivative is 
\begin{equation}
\partial_{\nu}\xi^{\mu}=\omega^{\mu}{}_{\nu},\qquad\partial_{\nu}\xi^{\nu}=0.\label{eq:flat-Poincare-Killing-derivative}
\end{equation}
For the curvature identity below, we choose 
\begin{equation}
F^{\mu\nu}[\xi]=\frac{1}{n-1}\left(x^{\mu}\xi^{\nu}-x^{\nu}\xi^{\mu}\right)+\frac{1}{2(n-1)}x^{2}\omega^{\mu\nu},\qquad x^{2}:=\eta_{\rho\sigma}x^{\rho}x^{\sigma}.\label{eq:flat-Poincare-Killing-potential}
\end{equation}
This representative satisfies 
\begin{equation}
\partial_{\nu}F^{\nu\mu}[\xi]=\xi^{\mu},\label{eq:flat-Poincare-potential-divergence}
\end{equation}
as can be checked easily. For a pure translation, $\omega_{\mu\nu}=0$,
and \eqref{eq:flat-Poincare-Killing-potential} reduces to the affine
potential 
\begin{equation}
F^{\mu\nu}[a]=\frac{1}{n-1}\left(x^{\mu}a^{\nu}-x^{\nu}a^{\mu}\right).\label{eq:flat-translation-potential}
\end{equation}
For a pure Lorentz generator, $a^{\mu}=0$, another simple solution
of \eqref{eq:flat-Killing-potential-definition} is 
\begin{equation}
F_{\mathrm{hom}}^{\mu\nu}[\omega]=-\frac{1}{2}x^{2}\omega^{\mu\nu}.\label{eq:flat-Lorentz-homogeneous-potential}
\end{equation}
The difference 
\begin{equation}
H^{\mu\nu}[\omega]:=F^{\mu\nu}[\omega]-F_{\mathrm{hom}}^{\mu\nu}[\omega]\label{eq:flat-Lorentz-potential-difference}
\end{equation}
is divergence-free: 
\begin{equation}
\partial_{\nu}H^{\nu\mu}[\omega]=0.\label{eq:flat-Lorentz-potential-difference-divergence}
\end{equation}
We use the representative \eqref{eq:flat-Poincare-Killing-potential},
rather than \eqref{eq:flat-Lorentz-homogeneous-potential}, because
its derivative is adapted to the algebraic Bianchi identity of the
$\mathcal{P}$-tensor. Lowering the indices with $\eta_{\mu\nu}$,
one finds 
\begin{align}
\partial_{\nu}F_{\beta\sigma}[\xi]=\frac{1}{n-1}\Bigl( & \eta_{\nu\beta}\xi_{\sigma}-\eta_{\nu\sigma}\xi_{\beta}+x_{\beta}\omega_{\sigma\nu}-x_{\sigma}\omega_{\beta\nu}+x_{\nu}\omega_{\beta\sigma}\Bigr).\label{eq:flat-Poincare-potential-derivative}
\end{align}
The last three terms can be written as 
\begin{equation}
x_{\beta}\omega_{\sigma\nu}-x_{\sigma}\omega_{\beta\nu}+x_{\nu}\omega_{\beta\sigma}=3x_{[\nu}\omega_{\beta\sigma]}.\label{eq:flat-potential-cyclic-part}
\end{equation}
This completely antisymmetric form is the reason for the particular
choice of representative in \eqref{eq:flat-Poincare-Killing-potential}.

The distinction from the AdS construction is therefore precise. In
AdS, the derivative of the Killing vector supplies the required two-form.
In flat space, the translational Killing two-form vanishes, and it
must be replaced by an antisymmetric potential whose divergence gives
the Killing vector.

\section{The flat-space curvature-flux identity}

\label{sec:flat-curvature-flux-identity}

We now linearize the exact identity 
\begin{equation}
\nabla_{\nu}\left(F_{\beta\sigma}\mathcal{P}^{\nu\mu\beta\sigma}\right)=\mathcal{P}^{\nu\mu\beta\sigma}\nabla_{\nu}F_{\beta\sigma}\label{eq:exact-PF-identity-flat-recall}
\end{equation}
about Minkowski spacetime. For $\Lambda=0$, the shifted tensor $\mathcal{P}^{\nu\mu\beta\sigma}$
coincides with the non-cosmological $\mathcal{P}$-tensor. Since the
background curvature and connection vanish in Cartesian coordinates,
\begin{equation}
\partial_{\nu}\left[F_{\beta\sigma}\left(\mathcal{P}^{\nu\mu\beta\sigma}\right)^{(1)}\right]=\left(\mathcal{P}^{\nu\mu\beta\sigma}\right)^{(1)}\partial_{\nu}F_{\beta\sigma}.\label{eq:flat-master-identity}
\end{equation}
We take $F_{\beta\sigma}$ to be the Poincaré Killing potential \eqref{eq:flat-Poincare-Killing-potential}.

The relevant traces of the linearized $\mathcal{P}$-tensor are 
\begin{align}
\eta_{\nu\beta}\left(\mathcal{P}^{\nu\mu\beta\sigma}\right)^{(1)}=-(n-3)\left(G^{\mu\sigma}\right)^{(1)},\qquad\eta_{\nu\sigma}\left(\mathcal{P}^{\nu\mu\beta\sigma}\right)^{(1)}=+(n-3)\left(G^{\mu\beta}\right)^{(1)}.\label{eq:flat-P-second-trace}
\end{align}
Consequently, the first two terms in \eqref{eq:flat-Poincare-potential-derivative}
give 
\begin{align}
 & \left(\mathcal{P}^{\nu\mu\beta\sigma}\right)^{(1)}\left(\eta_{\nu\beta}\xi_{\sigma}-\eta_{\nu\sigma}\xi_{\beta}\right)=-2(n-3)\xi_{\lambda}\left(G^{\mu\lambda}\right)^{(1)}.\label{eq:flat-P-trace-collapse-general}
\end{align}
The remaining part vanishes by the algebraic Bianchi identity. In
fact, using \eqref{eq:flat-potential-cyclic-part}, 
\begin{align}
 & \left(\mathcal{P}^{\nu\mu\beta\sigma}\right)^{(1)}\left(x_{\beta}\omega_{\sigma\nu}-x_{\sigma}\omega_{\beta\nu}+x_{\nu}\omega_{\beta\sigma}\right)=3\left(\mathcal{P}^{\nu\mu\beta\sigma}\right)^{(1)}x_{[\nu}\omega_{\beta\sigma]}=0,\label{eq:flat-Lorentz-dependent-contraction}
\end{align}
because 
\begin{equation}
\left(\mathcal{P}^{\nu\mu\beta\sigma}\right)^{(1)}+\left(\mathcal{P}^{\beta\mu\sigma\nu}\right)^{(1)}+\left(\mathcal{P}^{\sigma\mu\nu\beta}\right)^{(1)}=0.\label{eq:flat-P-algebraic-Bianchi}
\end{equation}
It follows that 
\begin{equation}
\left(\mathcal{P}^{\nu\mu\beta\sigma}\right)^{(1)}\partial_{\nu}F_{\beta\sigma}[\xi]=-\frac{2(n-3)}{n-1}\xi_{\lambda}\left(G^{\mu\lambda}\right)^{(1)}.\label{eq:flat-Poincare-collapse}
\end{equation}
Substitution into \eqref{eq:flat-master-identity} gives 
\begin{equation}
\xi_{\lambda}\left(G^{\mu\lambda}\right)^{(1)}=-\frac{n-1}{2(n-3)}\partial_{\nu}\left[F_{\beta\sigma}[\xi]\left(\mathcal{P}^{\nu\mu\beta\sigma}\right)^{(1)}\right].\label{eq:flat-Poincare-current-identity}
\end{equation}
This identity holds for every Poincaré Killing vector 
\begin{equation}
\xi^{\mu}=a^{\mu}+\omega^{\mu}{}_{\nu}x^{\nu}.\label{eq:flat-Poincare-Killing-recalled}
\end{equation}
The corresponding volume charge is 
\begin{equation}
Q[\xi]:=\frac{1}{\kappa_{n}}\int_{\Sigma}\dd\Sigma_{\mu}\,\xi_{\lambda}\left(G^{\mu\lambda}\right)^{(1)}.\label{eq:flat-Poincare-volume-charge}
\end{equation}
Using the Stokes' theorem, one obtains 
\begin{equation}
Q[\xi]=-\frac{n-1}{2\kappa_{n}(n-3)}\int_{\partial\Sigma}\dd\Sigma_{\mu\nu}\,F_{\beta\sigma}[\xi]\left(\mathcal{P}^{\nu\mu\beta\sigma}\right)^{(1)}.\label{eq:flat-Poincare-P-surface-charge-F}
\end{equation}
Inserting \eqref{eq:flat-Poincare-Killing-potential} gives 
\begin{align}
Q[\xi]=\frac{1}{2\kappa_{n}(n-3)}\int_{\partial\Sigma}\dd\Sigma_{\mu\nu}\,\left(\mathcal{P}^{\nu\mu\beta\sigma}\right)^{(1)}\Bigl( & \xi_{\beta}x_{\sigma}-\xi_{\sigma}x_{\beta}-\frac{1}{2}x^{2}\omega_{\beta\sigma}\Bigr).\label{eq:flat-Poincare-P-surface-charge}
\end{align}

If the asymptotic region is vacuum to the order relevant for the charge,
or more generally, if the corresponding Ricci and scalar-curvature
fluxes vanish, the $\mathcal{P}$-tensor may be replaced in the boundary
integral by the linearized Riemann tensor. Thus 
\begin{align}
Q[\xi]=\frac{1}{2\kappa_{n}(n-3)}\lim_{r\to\infty}\int_{S_{r}}\dd\Sigma_{\mu\nu}\,\left(R^{\nu\mu\beta\sigma}\right)^{(1)}\Bigl( & \xi_{\beta}x_{\sigma}-\xi_{\sigma}x_{\beta}-\frac{1}{2}x^{2}\omega_{\beta\sigma}\Bigr).\label{eq:flat-Poincare-Riemann-surface-charge}
\end{align}
For Lorentz generators, the weights in this expression grow quadratically
with $r$; the vanishing of the weighted Ricci contributions must
therefore be included among the asymptotic assumptions, together with
the Regge--Teitelboim parity conditions introduced below.

On an asymptotically Cartesian $t=\mathrm{constant}$ hypersurface,
the orientation 
\begin{equation}
\dd\Sigma_{\mu\nu}=n_{[\mu}\sigma_{\nu]}\dd S\label{eq:flat-binormal-orientation}
\end{equation}
implies 
\begin{equation}
\dd\Sigma_{\mu\nu}\left(R^{\nu\mu\beta\sigma}\right)^{(1)}=-\dd S_{i}\left(R^{i0\beta\sigma}\right)^{(1)},\label{eq:flat-binormal-Riemann-reduction}
\end{equation}
where we assumed $n^{\mu}$is the future-directed unit normal $n^{\mu}=(1,0,\ldots,0)$
and $\sigma^{\mu}=(0,\sigma^{i})$. Hence, the master formula takes
the spatial form

\noindent\fbox{\begin{minipage}[t]{1\columnwidth - 2\fboxsep - 2\fboxrule}%
\begin{align}
Q[\xi]=-\frac{1}{2\kappa_{n}(n-3)}\lim_{r\to\infty}\int_{S_{r}}\dd S_{i}\,\left(R^{i0\beta\sigma}\right)^{(1)}\Bigl( & \xi_{\beta}x_{\sigma}-\xi_{\sigma}x_{\beta}-\frac{1}{2}x^{2}\omega_{\beta\sigma}\Bigr).\label{eq:flat-Poincare-Riemann-spatial-charge}
\end{align}
\end{minipage}}\\
\\
The minus sign in \eqref{eq:flat-Poincare-Riemann-spatial-charge}
follows entirely from the chosen orientation and the reversed
index order $\nu\mu$ in the curvature tensor. It must be retained
in all component calculations.

With these conventions, the translation components will be identified
as 
\begin{equation}
E:=Q[\partial_{t}],\qquad P_{k}:=-Q[\partial_{k}].\label{eq:translation-charge-identification}
\end{equation}
The signs of the rotation and boost generators will be fixed explicitly
when those Killing vectors are introduced. For pure translations,
$\omega_{\mu\nu}=0$, while for pure Lorentz transformations $a^{\mu}=0$.
We now analyze these sectors separately. However, before that, let us make the following remark: in \cite{Penrose}, Penrose considered the type of conserved quantities built from the contraction of the Riemann tensor with the conformal Killing-Kano tensors ($K$) which are of the form $R_{\mu\nu\alpha\beta} K^{\alpha \beta}$. See \cite{Hull,Hull2} and the references therein for a detailed discussion and extension of the Penrose-type constructions. Further work is needed to establish the precise connections between our work here and these works. In our work, the conserved charge is directly built from Killing vectors from the start, and the antisymmetric potentials that appear on the codimension-2 surface are included; the connection of our charges to the ADM charges is clear. \footnote{We thank the authors of \cite{Hull} for reminding us of these works. }

\section{ADM four-momentum as a curvature flux}

\label{sec:ADM-four-momentum-curvature-flux}

We now specialize the flat-space master formula to translations. Let
$x^{\mu}=(t,x^{i})$ be asymptotically Cartesian coordinates, and
let the spacelike hypersurface $\Sigma$ approach a $t=\mathrm{constant}$
slice at spatial infinity. Its future-directed unit normal has the
asymptotic form 
\begin{equation}
n^{\mu}=(1,0,\ldots,0),\qquad n_{\mu}=(-1,0,\ldots,0).\label{eq:flat-timelike-normal}
\end{equation}
For large $r$, define 
\begin{equation}
r:=\left(\delta_{ij}x^{i}x^{j}\right)^{1/2},\qquad S_{r}:=\Sigma\cap\{r=\mathrm{constant}\}.\label{eq:Sr-definition}
\end{equation}
The outward-pointing unit normal to $S_{r}$ within $\Sigma$ is asymptotically
\begin{equation}
\sigma^{\mu}=(0,\sigma^{i}),\qquad\sigma^{i}=\frac{x^{i}}{r},\qquad\dd S_{i}:=\sigma_{i}\,\dd S.\label{eq:flat-spatial-area-element}
\end{equation}
With the binormal convention fixed before (\ref{eq:flat-binormal-orientation}),
one has 
\begin{equation}
\dd\Sigma_{\mu\nu}\left(R^{\nu\mu\beta\sigma}\right)^{(1)}=-\dd S_{i}\left(R^{i0\beta\sigma}\right)^{(1)}.\label{eq:flat-surface-orientation}
\end{equation}
The minus sign follows from $n_{0}=-1$ and from the reversed index
order $\nu\mu$ in the curvature tensor. It is fixed by the orientation
convention and will be used throughout the asymptotically flat analysis.

For a translation, 
\begin{equation}
\xi^{\mu}=a^{\mu},\qquad\omega_{\mu\nu}=0,\label{eq:flat-translation-sector}
\end{equation}
where $a^{\mu}$ is constant. Equation \eqref{eq:flat-Poincare-Riemann-spatial-charge}
reduces to 
\begin{equation}
Q[a]=-\frac{1}{\kappa_{n}(n-3)}\lim_{r\to\infty}\int_{S_{r}}\dd S_{i}\,\left(R^{i0\beta\sigma}\right)^{(1)}a_{\beta}x_{\sigma}.\label{eq:flat-translation-curvature-flux}
\end{equation}
We identify the ADM four-momentum through 
\begin{equation}
Q[a]=-a^{\mu}P_{\mu},\qquad P_{\mu}=(-E,P_{i}).\label{eq:flat-Poincare-pairing}
\end{equation}
Thus the future-directed unit time translation gives the energy, whereas
a unit spatial translation gives minus the corresponding covariant
spatial momentum component.

For the asymptotic time translation, 
\begin{equation}
a^{\mu}=(1,0,\ldots,0),\qquad a_{\mu}=(-1,0,\ldots,0),\label{eq:ADM-time-translation-vector}
\end{equation}
equation \eqref{eq:flat-translation-curvature-flux} gives

\noindent\fbox{\begin{minipage}[t]{1\columnwidth - 2\fboxsep - 2\fboxrule}%
\begin{equation}
E:=Q[\partial_{t}]=\frac{1}{\kappa_{n}(n-3)}\lim_{r\to\infty}\int_{S_{r}}\dd S_{i}\,x^{j}\left(R^{i0}{}_{j0}\right)^{(1)}.\label{eq:energy-curvature-main}
\end{equation}
\end{minipage}}\\
\\
Here we used 
\begin{equation}
R^{i0\,0j}=\eta^{00}\delta^{jk}R^{i0}{}_{0k}=R^{i0}{}_{j0},\label{eq:Riemann-index-relation-energy}
\end{equation}
where the second equality follows from antisymmetry in the last two
indices. The plus sign in \eqref{eq:energy-curvature-main} therefore
follows from the minus sign in \eqref{eq:flat-translation-curvature-flux}
together with $a_{0}=-1$.

For a spatial translation in the $k$-direction, 
\begin{equation}
a^{\mu}=\delta^{\mu}{}_{k},\label{eq:spatial-translation-vector-prelim}
\end{equation}
one has 
\begin{equation}
Q[\partial_{k}]=-\frac{1}{\kappa_{n}(n-3)}\lim_{r\to\infty}\int_{S_{r}}\dd S_{i}\,x^{j}\left(R^{i0}{}_{kj}\right)^{(1)}.\label{eq:spatial-translation-charge}
\end{equation}
Since $P_{k}=-Q[\partial_{k}]$, the spatial momentum is

\noindent\fbox{\begin{minipage}[t]{1\columnwidth - 2\fboxsep - 2\fboxrule}%
\begin{equation}
P_{k}:=-Q[\partial_{k}]=\frac{1}{\kappa_{n}(n-3)}\lim_{r\to\infty}\int_{S_{r}}\dd S_{i}\,x^{j}\left(R^{i0}{}_{kj}\right)^{(1)}.\label{eq:momentum-curvature-main}
\end{equation}
\end{minipage}}\\
\\
Equations~\eqref{eq:energy-curvature-main} and \eqref{eq:momentum-curvature-main}
are therefore the temporal and spatial components of the single translation
charge \eqref{eq:flat-translation-curvature-flux}.

We next show that \eqref{eq:energy-curvature-main} reproduces the
standard ADM energy. In the remainder of the asymptotically flat ADM
analysis, $h_{ij}$ denotes the perturbation of the induced spatial
metric. To avoid confusion with the spacetime trace introduced earlier,
define 
\begin{equation}
h_{i}^{\ i}:=\delta^{ij}h_{ij},\qquad V^{i}:=\partial_{j}h^{ij}-\partial^{i}h_{j}^{\ j}.\label{eq:ADM-vector}
\end{equation}
Spatial indices are raised and lowered with the asymptotic Euclidean
metric. With the normalization adopted in Section~\ref{sec:conventions},
the ADM energy is 
\begin{equation}
E_{{\rm ADM}}=\frac{1}{2\kappa_{n}}\lim_{r\to\infty}\int_{S_{r}}\dd S_{i}\,V^{i}.\label{eq:standard-ADM-energy}
\end{equation}
At linear order, the Gauss relation contains no term quadratic in
the extrinsic curvature. In the asymptotically vacuum region, the
spatial components of the spacetime Ricci tensor therefore give 
\begin{equation}
\left(R^{i0}{}_{j0}\right)^{(1)}=-\left(^{(n-1)}R^{i}{}_{j}\right)^{(1)}+o\left(r^{-(n-1)}\right).\label{eq:electric-Riemann-spatial-Ricci}
\end{equation}
The linearized Ricci tensor of the spatial metric satisfies 
\begin{equation}
2\left(^{(n-1)}R^{i}{}_{j}\right)^{(1)}=\partial_{j}V^{i}+\partial_{k}\left(\partial^{i}h^{k}{}_{j}-\partial^{k}h^{i}{}_{j}\right).\label{eq:linearized-spatial-Ricci-V}
\end{equation}
Multiplying by $x^{j}$, integrating derivatives by parts algebraically,
and using \eqref{eq:electric-Riemann-spatial-Ricci}, one finds 
\begin{equation}
x^{j}\left(R^{i0}{}_{j0}\right)^{(1)}=\frac{n-3}{2}V^{i}-\partial_{j}B^{ij}-\frac{1}{2}x^{i}\,{}^{(n-1)}R^{(1)}+o\left(r^{-(n-2)}\right),\label{eq:xR-to-ADM-local-correct}
\end{equation}
where 
\begin{equation}
^{(n-1)}R^{(1)}=\partial_{i}\partial_{j}h^{ij}-\partial^{2}h_{i}^{\ i}=\partial_{i}V^{i},\label{eq:linearized-spatial-scalar-curvature}
\end{equation}
and 
\begin{equation}
B^{ij}=\frac{1}{2}\left(x^{j}V^{i}-x^{i}V^{j}\right)-\frac{1}{2}x^{k}\left(\partial^{j}h^{i}{}_{k}-\partial^{i}h^{j}{}_{k}\right).\label{eq:Bij-explicit}
\end{equation}
By construction, 
\begin{equation}
B^{ij}=-B^{ji}.\label{eq:Bij-antisymmetry}
\end{equation}
For asymptotically vacuum initial data, the linearized Hamiltonian
constraint implies 
\begin{equation}
^{(n-1)}R^{(1)}=o\left(r^{-(n-1)}\right),\label{eq:asymptotic-Hamiltonian-constraint}
\end{equation}
and hence 
\begin{equation}
x^{i}\,{}^{(n-1)}R^{(1)}=o\left(r^{-(n-2)}\right).\label{eq:scalar-curvature-term-decay}
\end{equation}
The scalar-curvature term in \eqref{eq:xR-to-ADM-local-correct} therefore
gives no contribution to the charge. The antisymmetric improvement
also has zero flux through the closed surface $S_{r}$: 
\begin{equation}
\int_{S_{r}}\dd S_{i}\,\partial_{j}B^{ij}=0.\label{eq:Bij-no-surface-contribution}
\end{equation}
Indeed, the integrand is the pullback to $S_{r}$ of an exact $(n-2)$-form
constructed from $B^{ij}$, and its integral vanishes by Stokes' theorem
because $\partial S_{r}=\varnothing$. This argument requires only
that $B^{ij}$ be regular in a neighborhood of $S_{r}$; it does not
require regularity throughout the interior region. It follows that
\begin{equation}
\lim_{r\to\infty}\int_{S_{r}}\dd S_{i}\,x^{j}\left(R^{i0}{}_{j0}\right)^{(1)}=\frac{n-3}{2}\lim_{r\to\infty}\int_{S_{r}}\dd S_{i}\,V^{i}.\label{eq:curvature-to-ADM-energy-correct}
\end{equation}
Substituting this result into \eqref{eq:energy-curvature-main} gives
\begin{equation}
E=E_{{\rm ADM}}=\frac{1}{2\kappa_{n}}\lim_{r\to\infty}\int_{S_{r}}\dd S_{i}\,V^{i}.\label{eq:energy-equals-ADM-energy}
\end{equation}
Thus the curvature flux is not a new definition of the energy. It
is an equivalent curvature representative of the ADM energy. 

\subsection{Schwarzschild check}

\label{subsec:Schwarzschild-check-curvature-energy}

As a normalization and sign check, consider the four-dimensional Schwarzschild
metric in asymptotically Cartesian isotropic coordinates. To the order
relevant for the charge, 
\begin{equation}
g_{00}=-1+\frac{2GM}{r}+o(r^{-2}),\qquad g_{0i}=0,\qquad g_{ij}=\left(1+\frac{2GM}{r}\right)\delta_{ij}+o(r^{-2}).\label{eq:weak-Schwarzschild}
\end{equation}
Thus 
\begin{equation}
h_{00}=\frac{2GM}{r},\qquad h_{ij}=\frac{2GM}{r}\,\delta_{ij}.\label{eq:weak-Schwarzschild-perturbation}
\end{equation}
For a static perturbation of Minkowski spacetime, the curvature convention
\eqref{eq:Riemann-convention} gives 
\begin{equation}
\left(R^{i}{}_{0j0}\right)^{(1)}=-\frac{1}{2}\partial^{i}\partial_{j}h_{00}.\label{eq:Schwarzschild-Riemann-one-up}
\end{equation}
Using 
\begin{equation}
\partial^{i}\partial_{j}\frac{1}{r}=\frac{3\sigma^{i}\sigma_{j}-\delta^{i}{}_{j}}{r^{3}},\qquad\sigma^{i}:=\frac{x^{i}}{r},\label{eq:second-derivative-one-over-r}
\end{equation}
one obtains 
\begin{equation}
\left(R^{i}{}_{0j0}\right)^{(1)}=-GM\frac{3\sigma^{i}\sigma_{j}-\delta^{i}{}_{j}}{r^{3}}.\label{eq:Schwarzschild-Riemann-one-up-result}
\end{equation}
Raising the second index with $\eta^{00}=-1$ then gives 
\begin{equation}
\left(R^{i0}{}_{j0}\right)^{(1)}=-\left(R^{i}{}_{0j0}\right)^{(1)}=GM\frac{3\sigma^{i}\sigma_{j}-\delta^{i}{}_{j}}{r^{3}}.\label{eq:Schwarzschild-curvature-component}
\end{equation}
This explicit index-raising step is responsible for the positive sign
in the curvature component entering the energy formula. For the spatial
perturbation in \eqref{eq:weak-Schwarzschild-perturbation}, one has
\begin{equation}
h_{\mathrm{i}}^{\ i}=\frac{6GM}{r},\qquad V^{i}=\frac{4GM}{r^{2}}\sigma^{i}.\label{eq:Schwarzschild-ADM-vector}
\end{equation}
It follows directly that 
\begin{equation}
B^{ij}=0\label{eq:Schwarzschild-Bij-zero}
\end{equation}
in the asymptotic exterior region. Moreover, 
\begin{equation}
^{(3)}R^{(1)}=-4GM\,\partial^{2}\frac{1}{r}=0,\qquad r>0.\label{eq:Schwarzschild-spatial-scalar-zero}
\end{equation}
The last equality is an exterior-vacuum statement; distributionally,
$\partial^{2}(1/r)$ has support at the origin. Using 
\begin{equation}
\dd S_{i}=r^{2}\sigma_{i}\,\dd\Omega,\qquad x^{j}=r\sigma^{j},\label{eq:Schwarzschild-surface-elements}
\end{equation}
the curvature moment becomes 
\begin{align}
\int_{S_{r}}\dd S_{i}\,x^{j}\left(R^{i0}{}_{j0}\right)^{(1)} & =GM\int_{S^{2}}\dd\Omega\,\sigma_{i}\sigma^{j}\left(3\sigma^{i}\sigma_{j}-\delta^{i}{}_{j}\right)=2GM\int_{S^{2}}\dd\Omega=8\pi GM.\label{eq:Schwarzschild-curvature-integral}
\end{align}
Substitution into \eqref{eq:energy-curvature-main}, together with
$\kappa_{4}=8\pi G$, gives 
\begin{equation}
E=E_{{\rm ADM}}=M.\label{eq:Schwarzschild-energy-result}
\end{equation}
The Schwarzschild calculation therefore confirms both the normalization
and the sign of the curvature-flux energy formula.

\subsection{Spatial momentum}

\label{subsec:spatial-momentum-curvature-flux}

For a translation in the $k$-direction, 
\begin{equation}
a^{\mu}=\delta^{\mu}{}_{k},\label{eq:spatial-translation-vector}
\end{equation}
the translation charge gives 
\begin{equation}
P_{k}=\frac{1}{\kappa_{n}(n-3)}\lim_{r\to\infty}\int_{S_{r}}\dd S_{i}\,x^{j}\left(R^{i0}{}_{kj}\right)^{(1)}.\label{eq:momentum-curvature-main-repeat}
\end{equation}
Here we have used $P_{k}=-Q[\partial_{k}]$, consistently with \eqref{eq:flat-Poincare-pairing}.
We now reduce \eqref{eq:momentum-curvature-main-repeat} to the standard
ADM momentum. The extrinsic-curvature is 
\begin{equation}
K_{ij}:=-\gamma_{i}{}^{\mu}\gamma_{j}{}^{\nu}\nabla_{\mu}n_{\nu}.\label{eq:Kij-definition}
\end{equation}
For the ADM decomposition 
\begin{equation}
\dd s^{2}=-N^{2}\dd t^{2}+\gamma_{ij}\left(\dd x^{i}+N^{i}\dd t\right)\left(\dd x^{j}+N^{j}\dd t\right),\label{eq:ADM-decomposition}
\end{equation}
the future-directed unit normal is 
\begin{equation}
n^{\mu}=\frac{1}{N}\left(1,-N^{i}\right),\qquad n_{\mu}=\left(-N,0,\ldots,0\right),\label{eq:ADM-unit-normal}
\end{equation}
and 
\begin{equation}
K_{ij}=\frac{1}{2N}\left(D_{i}N_{j}+D_{j}N_{i}-\partial_{t}\gamma_{ij}\right).\label{eq:Kij-ADM-definition}
\end{equation}
Here $D_{i}$ is the covariant derivative compatible with the induced
metric, 
\begin{equation}
D_{i}\gamma_{jk}=0.\label{eq:spatial-metric-compatibility}
\end{equation}
Linearizing about a Cartesian slice of Minkowski spacetime, 
\begin{equation}
N=1+\order(h),\qquad N_{i}=h_{0i}+\order(h^{2}),\qquad\gamma_{ij}=\delta_{ij}+h_{ij},\label{eq:ADM-linear-identifications}
\end{equation}
gives 
\begin{equation}
\left(K_{ij}\right)^{(1)}=\frac{1}{2}\left(\partial_{i}h_{0j}+\partial_{j}h_{0i}-\partial_{0}h_{ij}\right).\label{eq:extrinsic-curvature-linear}
\end{equation}
We use 
\begin{equation}
\left(K^{i}{}_{j}\right)^{(1)}:=\delta^{i\ell}\left(K_{\ell j}\right)^{(1)},\qquad K^{(1)}:=\delta^{ij}\left(K_{ij}\right)^{(1)}.\label{eq:linear-K-index-conventions}
\end{equation}
With the curvature convention \eqref{eq:Riemann-convention} and the
extrinsic-curvature convention \eqref{eq:Kij-definition}, the linearized
Codazzi relation is 
\begin{equation}
\left(R_{0ijk}\right)^{(1)}=\partial_{j}\left(K_{ik}\right)^{(1)}-\partial_{k}\left(K_{ij}\right)^{(1)}.\label{eq:linear-Codazzi}
\end{equation}
Raising the first two indices with the Minkowski metric yields 
\begin{equation}
\left(R^{i0}{}_{kj}\right)^{(1)}=\partial_{k}\left(K^{i}{}_{j}\right)^{(1)}-\partial_{j}\left(K^{i}{}_{k}\right)^{(1)}.\label{eq:R-i0kj-Codazzi}
\end{equation}
It is useful to retain the linearized momentum-constraint residual
explicitly. Define 
\begin{equation}
\mathcal{M}_{k}^{(1)}:=\partial_{i}\left[\left(K^{i}{}_{k}\right)^{(1)}-K^{(1)}\delta^{i}{}_{k}\right].\label{eq:linear-momentum-constraint-residual}
\end{equation}
For asymptotically vacuum initial data, 
\begin{equation}
\mathcal{M}_{k}^{(1)}=o\left(r^{-(n-1)}\right),\label{eq:asymptotic-momentum-constraint}
\end{equation}
so that $x^{i}\mathcal{M}_{k}^{(1)}$ does not contribute to the limiting
surface integral. Using \eqref{eq:R-i0kj-Codazzi}, one has 
\begin{equation}
x^{j}\left(R^{i0}{}_{kj}\right)^{(1)}=x^{j}\left[\partial_{k}\left(K^{i}{}_{j}\right)^{(1)}-\partial_{j}\left(K^{i}{}_{k}\right)^{(1)}\right].\label{eq:xR-momentum-start}
\end{equation}
Introduce 
\begin{align}
C^{ij}{}_{k}:={} & x^{i}\left(K^{j}{}_{k}\right)^{(1)}-x^{j}\left(K^{i}{}_{k}\right)^{(1)}-\delta^{i}{}_{k}x_{\ell}\left(K^{j\ell}\right)^{(1)}+\delta^{j}{}_{k}x_{\ell}\left(K^{i\ell}\right)^{(1)}+\delta^{i}{}_{k}x^{j}K^{(1)}-\delta^{j}{}_{k}x^{i}K^{(1)}.\label{eq:Cijk-momentum-explicit}
\end{align}
This tensor is antisymmetric in its first two indices: 
\begin{equation}
C^{ij}{}_{k}=-C^{ji}{}_{k}.\label{eq:Cijk-antisymmetric}
\end{equation}
A direct differentiation gives 
\begin{align}
\partial_{j}C^{ij}{}_{k}={} & x^{j}\left[\partial_{k}\left(K^{i}{}_{j}\right)^{(1)}-\partial_{j}\left(K^{i}{}_{k}\right)^{(1)}\right]-(n-3)\left[\left(K^{i}{}_{k}\right)^{(1)}-K^{(1)}\delta^{i}{}_{k}\right]+x^{i}\mathcal{M}_{k}^{(1)}-\delta^{i}{}_{k}x^{\ell}\mathcal{M}_{\ell}^{(1)}.\label{eq:Cijk-divergence-check}
\end{align}
Consequently, 
\begin{align}
x^{j}\left(R^{i0}{}_{kj}\right)^{(1)}={} & (n-3)\left[\left(K^{i}{}_{k}\right)^{(1)}-K^{(1)}\delta^{i}{}_{k}\right]+\partial_{j}C^{ij}{}_{k}-x^{i}\mathcal{M}_{k}^{(1)}+\delta^{i}{}_{k}x^{\ell}\mathcal{M}_{\ell}^{(1)}.\label{eq:xR-to-ADM-momentum-exact}
\end{align}
Under the asymptotic condition \eqref{eq:asymptotic-momentum-constraint},
this reduces to 
\begin{equation}
x^{j}\left(R^{i0}{}_{kj}\right)^{(1)}=(n-3)\left[\left(K^{i}{}_{k}\right)^{(1)}-K^{(1)}\delta^{i}{}_{k}\right]+\partial_{j}C^{ij}{}_{k}+o\left(r^{-(n-2)}\right).\label{eq:xR-to-ADM-momentum-local}
\end{equation}
The antisymmetric improvement has vanishing flux through the closed
surface $S_{r}$: 
\begin{equation}
\int_{S_{r}}\dd S_{i}\,\partial_{j}C^{ij}{}_{k}=0.\label{eq:Cijk-no-contribution}
\end{equation}
Equivalently, after dualizing $C^{ij}{}_{k}$ in its antisymmetric
pair, the integrand is the pullback of an exact $(n-2)$-form to $S_{r}$;
its integral therefore vanishes because $\partial S_{r}=\varnothing$.
Only regularity in a neighborhood of the asymptotic surface is required.
It follows that 
\begin{align}
\lim_{r\to\infty}\int_{S_{r}}\dd S_{i}\,x^{j}\left(R^{i0}{}_{kj}\right)^{(1)}=(n-3)\lim_{r\to\infty}\int_{S_{r}}\dd S_{i}\left[\left(K^{i}{}_{k}\right)^{(1)}-K^{(1)}\delta^{i}{}_{k}\right].\label{eq:curvature-to-ADM-momentum}
\end{align}
Since only the leading asymptotic part of $K_{ij}$ contributes, the
linearized quantities may equivalently be replaced by the full extrinsic
curvature: 
\begin{align}
\lim_{r\to\infty}\int_{S_{r}}\dd S_{i}\,x^{j}\left(R^{i0}{}_{kj}\right)^{(1)}=(n-3)\lim_{r\to\infty}\int_{S_{r}}\dd S_{i}\left(K^{i}{}_{k}-K\delta^{i}{}_{k}\right).\label{eq:curvature-to-ADM-momentum-full-K}
\end{align}
Substitution into \eqref{eq:momentum-curvature-main-repeat} gives
\begin{equation}
P_{k}=\frac{1}{\kappa_{n}}\lim_{r\to\infty}\int_{S_{r}}\dd S_{i}\left(K^{i}{}_{k}-K\delta^{i}{}_{k}\right).\label{eq:standard-ADM-momentum}
\end{equation}
This is the standard ADM spatial momentum with the sign determined
by the extrinsic-curvature convention \eqref{eq:Kij-definition}.

\subsubsection{Boosted Schwarzschild check}

\label{subsec:boosted-Schwarzschild-momentum-check}

As a direct check of the momentum formula, consider the weak field
of a four-dimensional Schwarzschild black hole of rest mass $M$,
moving in the positive $z$-direction with constant velocity $v$.
We keep all orders in $v$, but retain only the terms linear in $GM$.
Define 
\begin{equation}
\gamma:=\frac{1}{\sqrt{1-v^{2}}}.\label{eq:gamma-def}
\end{equation}
In the rest-frame coordinates $(T,X,Y,Z)$, the asymptotic Schwarzschild
metric in isotropic form is 
\begin{equation}
\dd s^{2}=-\left(1-\frac{2GM}{\rho}\right)\dd T^{2}+\left(1+\frac{2GM}{\rho}\right)\left(\dd X^{2}+\dd Y^{2}+\dd Z^{2}\right)+\order(\rho^{-2}),\label{eq:rest-schwarzschild-weak}
\end{equation}
where 
\begin{equation}
\rho:=\sqrt{X^{2}+Y^{2}+Z^{2}}.\label{eq:rho-rest-frame}
\end{equation}
The Lorentz transformation to the asymptotic coordinates $(t,x,y,z)$
is 
\begin{equation}
T=\gamma(t-vz),\qquad Z=\gamma(z-vt),\qquad X=x,\qquad Y=y.\label{eq:boost-coordinates}
\end{equation}
Consequently, 
\begin{equation}
\rho(t)=\sqrt{x^{2}+y^{2}+\gamma^{2}(z-vt)^{2}}.\label{eq:rho-time-dependent}
\end{equation}
It is important that the derivatives entering the curvature and the
extrinsic curvature be evaluated using the full time-dependent function
$\rho(t)$, before restricting to the hypersurface $t=0$. Define
\begin{equation}
\Phi(t):=\frac{2GM}{\rho(t)}.\label{eq:Phi-def}
\end{equation}
The nonvanishing components of the boosted perturbation are 
\begin{align}
h_{00} & =\gamma^{2}(1+v^{2})\Phi(t), & h_{03} & =-2\gamma^{2}v\,\Phi(t),\nonumber \\
h_{33} & =\gamma^{2}(1+v^{2})\Phi(t), & h_{11} & =h_{22}=\Phi(t).\label{eq:h-boosted-components}
\end{align}
On the hypersurface $t=0$, let 
\begin{equation}
\rho_{0}:=\rho(0)=\sqrt{x^{2}+y^{2}+\gamma^{2}z^{2}}.\label{eq:rho-boosted-slice}
\end{equation}
The linearized Riemann tensor with all indices lowered is 
\begin{align}
\left(R_{\alpha\beta\mu\nu}\right)^{(1)}=\frac{1}{2}\Bigl( & \partial_{\mu}\partial_{\beta}h_{\alpha\nu}+\partial_{\nu}\partial_{\alpha}h_{\beta\mu}-\partial_{\nu}\partial_{\beta}h_{\alpha\mu}-\partial_{\mu}\partial_{\alpha}h_{\beta\nu}\Bigr).\label{eq:boost-linearized-riemann}
\end{align}
By reflection symmetry in the $x$- and $y$-directions, only the
$z$-component of the momentum can be nonzero. On a large coordinate
sphere, write 
\begin{equation}
x^{i}=r\sigma^{i},\qquad\sigma^{i}\sigma_{i}=1,\qquad\dd S_{i}=r^{2}\sigma_{i}\,\dd\Omega,\qquad u:=\sigma^{3}=\cos\theta.\label{eq:sphere-data-boosted}
\end{equation}
It is convenient to introduce 
\begin{equation}
\Delta(u):=1+(\gamma^{2}-1)u^{2}.\label{eq:boost-angular-Delta}
\end{equation}
Then 
\begin{equation}
\rho_{0}=r\sqrt{\Delta(u)}.\label{eq:rho-asymptotic-angular}
\end{equation}
A direct evaluation of the curvature component gives 
\begin{equation}
r^{3}\sigma_{i}\sigma^{j}\left(R^{i0}{}_{3j}\right)^{(1)}=3GM\gamma^{2}v\,\frac{1-u^{2}}{\Delta(u)^{5/2}}.\label{eq:angular-integrand}
\end{equation}
Since 
\begin{equation}
\dd S_{i}\,x^{j}\left(R^{i0}{}_{3j}\right)^{(1)}=r^{3}\sigma_{i}\sigma^{j}\left(R^{i0}{}_{3j}\right)^{(1)}\dd\Omega,\label{eq:boosted-flux-angular-reduction}
\end{equation}
the required angular integral is 
\begin{align}
\int_{S^{2}}\frac{1-u^{2}}{\Delta(u)^{5/2}}\,\dd\Omega & =2\pi\int_{-1}^{1}\frac{1-u^{2}}{\left[1+(\gamma^{2}-1)u^{2}\right]^{5/2}}\,\dd u=\frac{8\pi}{3\gamma}.\label{eq:boost-angular-integral}
\end{align}
Therefore, 
\begin{equation}
\lim_{r\to\infty}\int_{S_{r}}\dd S_{i}\,x^{j}\left(R^{i0}{}_{3j}\right)^{(1)}=8\pi G\,\gamma Mv.\label{eq:boosted-curvature-flux-result}
\end{equation}
Using $\kappa_{4}=8\pi G$, equation \eqref{eq:momentum-curvature-main-repeat}
gives 
\begin{equation}
P_{3}=\gamma Mv,\qquad P_{1}=P_{2}=0.\label{eq:P3-result}
\end{equation}
Rotating the direction of motion gives 
\begin{equation}
P_{i}=\gamma Mv_{i}.\label{eq:boosted-spatial-momentum-general}
\end{equation}
A parallel evaluation of the time-translation charge gives 
\begin{equation}
E=\gamma M.\label{eq:boosted-energy-result}
\end{equation}
Equivalently, this follows from the asymptotic Lorentz covariance
of the translation charges. Thus 
\begin{equation}
P^{\mu}=(E,P^{i})=\gamma M(1,v^{i}).\label{eq:boosted-schwarzschild-four-momentum}
\end{equation}
We may verify the same result directly from the ADM momentum expression.
Using \eqref{eq:extrinsic-curvature-linear}, with the time derivatives
taken before setting $t=0$, one finds that the trace $K$ has a nonvanishing
term of order $r^{-2}$. The relevant surface density is therefore
the full combination $K^{i}{}_{k}-K\delta^{i}{}_{k}$, rather than
$K^{i}{}_{k}$ alone. For $k=3$, 
\begin{equation}
r^{2}\sigma_{i}\left(K^{i}{}_{3}-K\delta^{i}{}_{3}\right)=\frac{2GM\gamma^{2}v}{\Delta(u)^{3/2}}+\order(r^{-1}).\label{eq:boosted-schwarzschild-Kij}
\end{equation}
For completeness, the leading trace itself is 
\begin{equation}
K=\frac{GM\gamma^{2}v(2\gamma^{2}-1)u}{r^{2}\Delta(u)^{3/2}}+\order(r^{-3}).\label{eq:boosted-Schwarzschild-K-trace}
\end{equation}
The remaining angular integral is 
\begin{align}
\int_{S^{2}}\frac{\dd\Omega}{\Delta(u)^{3/2}} & =2\pi\int_{-1}^{1}\frac{\dd u}{\left[1+(\gamma^{2}-1)u^{2}\right]^{3/2}}=\frac{4\pi}{\gamma}.\label{eq:boosted-ADM-angular-integral}
\end{align}
Hence 
\begin{align}
P_{3} & =\frac{1}{8\pi G}\lim_{r\to\infty}\int_{S_{r}}\dd S_{i}\left(K^{i}{}_{3}-K\delta^{i}{}_{3}\right)=\frac{1}{8\pi G}\left(2GM\gamma^{2}v\right)\frac{4\pi}{\gamma}=\gamma Mv.\label{eq:Pk-final-extrinsic}
\end{align}
The curvature flux and the standard ADM surface integral therefore
give the same momentum, 
\begin{equation}
P_{i}=\gamma Mv_{i}.\label{eq:boosted-final-momentum}
\end{equation}

\section{Lorentz charges and Regge--Teitelboim completion}

\label{sec:Lorentz-charges-RT-completion}

We now turn to the Lorentz sector of the asymptotically flat charges.
In four spacetime dimensions, Lorentz Killing vectors grow linearly
in the asymptotically Cartesian coordinates. Their surface integrals
are therefore more delicate than the translation charges: the leading
terms are potentially divergent, and additional parity conditions
are required to obtain finite and unambiguous canonical generators.

We impose the standard Regge--Teitelboim asymptotic conditions \cite{ReggeTeitelboim,BeigOMurchadha}.
In asymptotically Cartesian coordinates, the spatial metric has the
expansion 
\begin{equation}
\gamma_{ij}=\delta_{ij}+\frac{h_{ij}^{(1)}(\hat{x})}{r}+\frac{h_{ij}^{(2)}(\hat{x})}{r^{2}}+o(r^{-2}),\label{eq:RT-parity-metric}
\end{equation}
with 
\begin{equation}
h_{ij}^{(1)}(-\hat{x})=h_{ij}^{(1)}(\hat{x}).\label{eq:RT-leading-metric-even}
\end{equation}
Thus the leading spatial metric perturbation is even under the antipodal
map $\hat{x}^{i}\mapsto-\hat{x}^{i}$. It is convenient to introduce
the ADM momentum combination 
\begin{equation}
\Pi^{i}{}_{j}:=K^{i}{}_{j}-K\delta^{i}{}_{j}.\label{eq:Pi-ADM-definition}
\end{equation}
Its asymptotic expansion is taken to be 
\begin{equation}
\Pi^{i}{}_{j}=\frac{\Pi^{(2)i}{}_{j}(\hat{x})}{r^{2}}+\frac{\Pi^{(3)i}{}_{j}(\hat{x})}{r^{3}}+o(r^{-3}),\label{eq:RT-parity-momentum}
\end{equation}
where 
\begin{equation}
\Pi^{(2)i}{}_{j}(-\hat{x})=-\Pi^{(2)i}{}_{j}(\hat{x}).\label{eq:RT-leading-momentum-odd}
\end{equation}
No parity restriction is imposed here on the subleading coefficients
$h_{ij}^{(2)}$ and $\Pi^{(3)i}{}_{j}$. These subleading terms contain,
in particular, the finite center-of-mass and angular-momentum data.

The parity conditions have a transparent role. Since a Lorentz Killing
vector grows as $r$, the leading $r^{-2}$ term in $\Pi^{i}{}_{j}$
would apparently produce a linearly divergent surface integral. Its
angular coefficient, however, is odd under the antipodal map and therefore
integrates to zero. The next term, proportional to $r^{-3}$, gives
the finite Lorentz charge. This is why the $r^{-3}$ term in \eqref{eq:RT-parity-momentum}
must not be excluded.

For a Lorentz Killing vector, 
\begin{equation}
\xi^{\mu}[\omega]=\omega^{\mu}{}_{\nu}x^{\nu},\qquad\omega_{\mu\nu}=-\omega_{\nu\mu},\label{eq:Lorentz-Killing-vector-RT}
\end{equation}
the four-dimensional master formula \eqref{eq:flat-Poincare-Riemann-spatial-charge},
evaluated on the $t=0$ hypersurface where $x^{2}=r^{2}$, gives 
\begin{align}
Q[\omega]=-\frac{1}{2\kappa_{4}}\lim_{r\to\infty}\int_{S_{r}}\dd S_{i}\,\left(R^{i0\beta\sigma}\right)^{(1)}\Bigl( & \xi_{\beta}x_{\sigma}-\xi_{\sigma}x_{\beta}-\frac{1}{2}r^{2}\omega_{\beta\sigma}\Bigr).\label{eq:Lorentz-charge-t-zero}
\end{align}
The final term is the contribution of the Bianchi-adapted Killing
potential. It is essential to the flat-space current identity and
must be retained in both the rotation and boost sectors.

\subsection{Rotations}

\label{subsec:rotations-curvature-flux}

Consider the rotation Killing vector in the $(k,l)$-plane, 
\begin{equation}
\xi_{(kl)}=x_{k}\frac{\partial}{\partial x^{l}}-x_{l}\frac{\partial}{\partial x^{k}}.\label{eq:rotation-killing-vector}
\end{equation}
With the Poincaré pairing convention adopted above, 
\begin{equation}
J_{kl}=-Q[\xi_{(kl)}].\label{eq:rotation-charge-pairing}
\end{equation}
Equation~\eqref{eq:Lorentz-charge-t-zero} then gives 
\begin{equation}
J_{kl}^{{\rm curv}}=\frac{1}{\kappa_{4}}\lim_{r\to\infty}\int_{S_{r}}\dd S_{i}\,\mathcal{J}^{i}{}_{kl},\label{eq:angular-momentum-curvature-flux}
\end{equation}
where 
\begin{equation}
\mathcal{J}^{i}{}_{kl}:=x_{k}x^{j}\left(R^{i0}{}_{lj}\right)^{(1)}-x_{l}x^{j}\left(R^{i0}{}_{kj}\right)^{(1)}+\frac{1}{2}r^{2}\left(R^{i0}{}_{kl}\right)^{(1)}.\label{eq:angular-curvature-density}
\end{equation}
The last term in \eqref{eq:angular-curvature-density} comes entirely
from the $-\frac{1}{2}x^{2}\omega_{\beta\sigma}$ term in the Killing
potential.

Since the curvature expression is linearized, in the following local
manipulations $K_{ij}$, $K$, and $\Pi^{i}{}_{j}$ denote their leading
linearized asymptotic parts. The superscript $(1)$ is suppressed
to avoid cluttering the formulas. The linearized Codazzi identity
is 
\begin{equation}
\left(R^{i0}{}_{kj}\right)^{(1)}=\partial_{k}K^{i}{}_{j}-\partial_{j}K^{i}{}_{k}.\label{eq:linearized-codazzi-angular}
\end{equation}
Define the linearized momentum-constraint residual by 
\begin{equation}
\mathcal{M}_{k}:=\partial_{i}\Pi^{i}{}_{k}=\partial_{i}\left(K^{i}{}_{k}-K\delta^{i}{}_{k}\right).\label{eq:momentum-constraint-residual-angular}
\end{equation}
We first recall the antisymmetric tensor (\ref{eq:Cijk-momentum-explicit})
and introduce 
\begin{align}
\mathcal{U}^{ij}{}_{kl}:={} & x_{k}C^{ij}{}_{l}-x_{l}C^{ij}{}_{k}+\frac{1}{2}r^{2}\left(\delta^{j}{}_{k}K^{i}{}_{l}-\delta^{j}{}_{l}K^{i}{}_{k}-\delta^{i}{}_{k}K^{j}{}_{l}+\delta^{i}{}_{l}K^{j}{}_{k}\right)-\frac{1}{2}r^{2}K\left(\delta^{i}{}_{l}\delta^{j}{}_{k}-\delta^{i}{}_{k}\delta^{j}{}_{l}\right).\label{eq:U-angular-explicit}
\end{align}
It is antisymmetric in its first two indices: 
\begin{equation}
\mathcal{U}^{ij}{}_{kl}=-\mathcal{U}^{ji}{}_{kl}.\label{eq:U-angular-antisymmetric}
\end{equation}
A direct differentiation, using \eqref{eq:linearized-codazzi-angular},
gives the exact linearized identity 
\begin{align}
\mathcal{J}^{i}{}_{kl}={} & x_{k}\Pi^{i}{}_{l}-x_{l}\Pi^{i}{}_{k}+\partial_{j}\mathcal{U}^{ij}{}_{kl}+\mathcal{R}^{i}{}_{kl}[\mathcal{M}],\label{eq:angular-curvature-to-ADM-exact}
\end{align}
where the momentum-constraint remainder is 
\begin{align}
\mathcal{R}^{i}{}_{kl}[\mathcal{M}]:={} & -x_{k}x^{i}\mathcal{M}_{l}+x_{l}x^{i}\mathcal{M}_{k}+\left(x_{k}\delta^{i}{}_{l}-x_{l}\delta^{i}{}_{k}\right)x^{m}\mathcal{M}_{m}+\frac{1}{2}r^{2}\left(\delta^{i}{}_{k}\mathcal{M}_{l}-\delta^{i}{}_{l}\mathcal{M}_{k}\right).\label{eq:angular-momentum-constraint-remainder}
\end{align}
Thus the reduction to the canonical angular momentum is not purely
kinematical: it also uses the asymptotic momentum constraint.

For asymptotically vacuum initial data, we assume the momentum constraint
in the weighted form required by the Lorentz charge, 
\begin{equation}
\lim_{r\to\infty}\int_{S_{r}}\dd S_{i}\,\mathcal{R}^{i}{}_{kl}[\mathcal{M}]=0.\label{eq:weighted-momentum-constraint-rotation}
\end{equation}
Equation~\eqref{eq:angular-curvature-to-ADM-exact} then becomes,
in the asymptotic sense relevant to the charge, 
\begin{equation}
\mathcal{J}^{i}{}_{kl}=x_{k}\Pi^{i}{}_{l}-x_{l}\Pi^{i}{}_{k}+\partial_{j}\mathcal{U}^{ij}{}_{kl}+o(r^{-2}).\label{eq:angular-curvature-to-ADM-local}
\end{equation}

The antisymmetric improvement has vanishing flux through the closed
surface $S_{r}$: 
\begin{equation}
\int_{S_{r}}\dd S_{i}\,\partial_{j}\mathcal{U}^{ij}{}_{kl}=0.\label{eq:U-angular-no-contribution}
\end{equation}
Indeed, after dualizing the antisymmetric pair $i,j$, the integrand
is the pullback of an exact one-form to $S_{r}$. Its integral vanishes
by Stokes' theorem because $\partial S_{r}=\varnothing$. Only regularity
in a neighborhood of the asymptotic sphere is required; no assumption
about the black-hole interior is involved. It follows that 
\begin{equation}
J_{kl}^{{\rm curv}}=\frac{1}{\kappa_{4}}\lim_{r\to\infty}\int_{S_{r}}\dd S_{i}\left(x_{k}\Pi^{i}{}_{l}-x_{l}\Pi^{i}{}_{k}\right).\label{eq:curvature-equals-ADM-angular-Pi}
\end{equation}
Restoring the full asymptotic extrinsic curvature, 
\begin{align}
J_{kl}^{{\rm curv}}=\frac{1}{\kappa_{4}}\lim_{r\to\infty}\int_{S_{r}}\dd S_{i}\Bigl[ & x_{k}\left(K^{i}{}_{l}-K\delta^{i}{}_{l}\right)-x_{l}\left(K^{i}{}_{k}-K\delta^{i}{}_{k}\right)\Bigr].\label{eq:curvature-equals-ADM-angular}
\end{align}
This is precisely the Regge--Teitelboim/ADM angular momentum: 
\begin{equation}
J_{kl}^{{\rm curv}}=J_{kl}^{{\rm ADM}}.\label{eq:curvature-ADM-angular-equality}
\end{equation}
The role of the parity conditions can now be seen directly. The leading
term in $\Pi^{i}{}_{j}$ is odd, whereas $x_{k}$ and $\dd S_{i}$
are both odd under the antipodal map. The potentially divergent leading
integrand is therefore odd and integrates to zero. The finite angular
momentum is determined by the subleading $r^{-3}$ part of $\Pi^{i}{}_{j}$.

\subsection{Boosts and the center of mass}

\label{subsec:boosts-center-of-mass-curvature-flux}

We now consider the Lorentz boosts. For a boost in the $k$-direction,
we use the Minkowski Killing vector 
\begin{equation}
\xi_{(0k)}=x_{k}\frac{\partial}{\partial t}+t\frac{\partial}{\partial x^{k}}.\label{eq:boost-xi}
\end{equation}
For earlier discussions of the relation between asymptotic boost generators
and the gravitational center of mass, see ~\cite{Petrov1,Petrov2}
and the references therein. In the notation 
\begin{equation}
\xi^{\mu}=\omega^{\mu}{}_{\nu}x^{\nu},\label{eq:boost-omega-definition}
\end{equation}
the nonvanishing mixed components are 
\begin{equation}
\omega^{0}{}_{k}=\omega^{k}{}_{0}=1,\label{eq:boost-omega-mixed-components}
\end{equation}
whereas the corresponding lowered components satisfy 
\begin{equation}
\omega_{0k}=-1,\qquad\omega_{k0}=1.\label{eq:boost-omega-lowered-components}
\end{equation}
Thus $\omega_{\mu\nu}=-\omega_{\nu\mu}$, as required. On the hypersurface
$t=0$, 
\begin{equation}
\xi_{(0k)}^{0}=x_{k},\qquad\xi_{(0k)}^{i}=0,\qquad x^{2}=r^{2}.\label{eq:boost-vector-t-zero}
\end{equation}
We define the corresponding conserved boost charge by 
\begin{equation}
\mathcal{K}_{k}:=Q[\xi_{(0k)}].\label{eq:boost-charge-definition}
\end{equation}
Substitution of \eqref{eq:boost-vector-t-zero} and \eqref{eq:boost-omega-lowered-components}
into the Lorentz curvature-flux formula \eqref{eq:Lorentz-charge-t-zero}
gives 
\begin{equation}
\mathcal{K}_{k}^{{\rm curv}}=\frac{1}{\kappa_{4}}\lim_{r\to\infty}\int_{S_{r}}\dd S_{i}\,\mathcal{B}^{i}{}_{k},\label{eq:boost-curvature-flux}
\end{equation}
where 
\begin{equation}
\mathcal{B}^{i}{}_{k}:=x_{k}x^{j}\left(R^{i0}{}_{j0}\right)^{(1)}-\frac{1}{2}r^{2}\left(R^{i0}{}_{k0}\right)^{(1)}.\label{eq:boost-curvature-density}
\end{equation}
The first term is the first spatial moment of the electric curvature
flux. The second term originates from the $-x^{2}\omega_{\beta\sigma}/2$
part of the Bianchi-adapted Killing potential. We shall see that this
second term is precisely what generates the metric subtraction required
in the Regge--Teitelboim boost charge.

\subsubsection{Reduction to spatial curvature}

We have introduced
\begin{equation}
V^{i}:=\partial_{j}h^{ij}-\partial^{i}h_{j}^{j}.\label{eq:RT-boost-Vi}
\end{equation}
We also introduce 
\begin{equation}
W^{ij}{}_{k}:=\partial^{i}h^{j}{}_{k}-\partial^{j}h^{i}{}_{k},\qquad W^{ij}{}_{k}=-W^{ji}{}_{k}.\label{eq:boost-W-definition}
\end{equation}
The linearized Ricci tensor and scalar curvature of the spatial metric
are 
\begin{align}
2\mathscr{R}^{i}{}_{k} & :=2\left(^{(3)}R^{i}{}_{k}\right)^{(1)}=\partial_{k}V^{i}+\partial_{j}W^{ij}{}_{k},\label{eq:linearized-spatial-Ricci-decomposition}\\
\mathscr{R} & :=\left(^{(3)}R\right)^{(1)}=\partial_{i}V^{i}.\label{eq:linearized-Hamiltonian-constraint-boost}
\end{align}
For completeness, the first identity follows from 
\begin{align}
2\left(^{(3)}R^{i}{}_{k}\right)^{(1)}={} & \partial_{j}\partial^{i}h^{j}{}_{k}+\partial_{j}\partial_{k}h^{ij}-\partial^{2}h^{i}{}_{k}-\partial^{i}\partial_{k}h_{j}^{j}\nonumber \\
={} & \partial_{k}\left(\partial_{j}h^{ij}-\partial^{i}h_{j}^{j}\right)+\partial_{j}\left(\partial^{i}h^{j}{}_{k}-\partial^{j}h^{i}{}_{k}\right).\label{eq:linearized-spatial-Ricci-derivation}
\end{align}
The contracted Gauss relation fixes the sign connecting the electric
spacetime curvature to the spatial Ricci tensor. At linear order,
the terms quadratic in the extrinsic curvature vanish, and one obtains
\begin{equation}
\left(R^{i0}{}_{k0}\right)^{(1)}=\left(R^{i}{}_{k}\right)^{(1)}-\mathscr{R}^{i}{}_{k},\label{eq:contracted-Gauss-boost}
\end{equation}
where $(R^{i}{}_{k})^{(1)}$ on the right-hand side is the mixed spacetime
Ricci tensor. In an asymptotically vacuum region, 
\begin{equation}
\left(R^{i}{}_{k}\right)^{(1)}=o(r^{-4})\label{eq:asymptotic-spacetime-Ricci-boost}
\end{equation}
to the order relevant for the four-dimensional boost charge. Thus
\begin{equation}
\left(R^{i0}{}_{k0}\right)^{(1)}=-\mathscr{R}^{i}{}_{k}+o(r^{-4}).\label{eq:electric-to-spatial-Ricci}
\end{equation}
It is nevertheless useful to retain the spacetime Ricci remainder
explicitly. Define 
\begin{equation}
\mathscr{S}^{i}{}_{k}:=\left(R^{i}{}_{k}\right)^{(1)}.\label{eq:boost-spacetime-Ricci-residual}
\end{equation}
Equations~\eqref{eq:boost-curvature-density} and \eqref{eq:contracted-Gauss-boost}
then give 
\begin{align}
\mathcal{B}^{i}{}_{k}={} & -x_{k}x^{j}\mathscr{R}^{i}{}_{j}+\frac{1}{2}r^{2}\mathscr{R}^{i}{}_{k}+x_{k}x^{j}\mathscr{S}^{i}{}_{j}-\frac{1}{2}r^{2}\mathscr{S}^{i}{}_{k}.\label{eq:boost-density-Gauss-decomposition}
\end{align}
The first line is determined by the intrinsic geometry of the spatial
slice. The second line vanishes under the asymptotic vacuum field
equations, with the weighted falloff required by the boost integral.

\subsubsection{The central spatial identity}

The spatial-curvature combination appearing in the first line of \eqref{eq:boost-density-Gauss-decomposition}
admits an exact linear decomposition. Define 
\begin{equation}
U^{ij}:=\frac{1}{2}\left(x^{i}V^{j}-x^{j}V^{i}-x^{a}W^{ij}{}_{a}\right),\qquad U^{ij}=-U^{ji}.\label{eq:boost-U-improvement}
\end{equation}
Also define the antisymmetric tensor 
\begin{align}
\mathcal{Y}^{ij}{}_{k}:={} & x_{k}U^{ij}+\frac{1}{4}r^{2}W^{ij}{}_{k}+\frac{1}{4}r^{2}\left(\delta^{j}{}_{k}V^{i}-\delta^{i}{}_{k}V^{j}\right)-\frac{1}{2}\left(x^{i}h^{j}{}_{k}-x^{j}h^{i}{}_{k}\right)\nonumber \\
 & +\frac{1}{2}\left(x^{i}\delta^{j}{}_{k}-x^{j}\delta^{i}{}_{k}\right)h_{j}^{j}-\frac{1}{2}\left(\delta^{j}{}_{k}x^{a}h^{i}{}_{a}-\delta^{i}{}_{k}x^{a}h^{j}{}_{a}\right).\label{eq:boost-Y-improvement}
\end{align}
Every term on the right-hand side is antisymmetric in $i,j$, and
hence 
\begin{equation}
\mathcal{Y}^{ij}{}_{k}=-\mathcal{Y}^{ji}{}_{k}.\label{eq:boost-Y-antisymmetry}
\end{equation}
A direct use of \eqref{eq:linearized-spatial-Ricci-decomposition}
gives the exact identity 
\begin{align}
-x_{k}x^{j}\mathscr{R}^{i}{}_{j}+\frac{1}{2}r^{2}\mathscr{R}^{i}{}_{k}={} & \frac{1}{2}\left[x_{k}V^{i}-\left(h^{i}{}_{k}-\delta^{i}{}_{k}h_{j}^{j}\right)\right]-\frac{1}{2}x_{k}x^{i}\mathscr{R}+\frac{1}{4}\delta^{i}{}_{k}r^{2}\mathscr{R}+\partial_{j}\mathcal{Y}^{ij}{}_{k}.\label{eq:central-boost-spatial-identity}
\end{align}
This is the central identity in the boost calculation. It separates
the curvature moment into four conceptually distinct pieces:
\begin{enumerate}
\item the Regge--Teitelboim surface density; 
\item terms proportional to the linearized Hamiltonian constraint $\mathscr{R}$; 
\item the divergence of an antisymmetric improvement; 
\item through \eqref{eq:boost-density-Gauss-decomposition}, terms proportional
to the asymptotic spacetime field equations. 
\end{enumerate}
Combining \eqref{eq:boost-density-Gauss-decomposition} and \eqref{eq:central-boost-spatial-identity},
we obtain 
\begin{align}
\mathcal{B}^{i}{}_{k}={} & \frac{1}{2}\left[x_{k}\left(\partial_{j}h^{ij}-\partial^{i}h_{j}^{j}\right)-\left(h^{i}{}_{k}-\delta^{i}{}_{k}h_{j}^{j}\right)\right]+\partial_{j}\mathcal{Y}^{ij}{}_{k}+\mathcal{H}^{i}{}_{k}+\mathcal{E}^{i}{}_{k},\label{eq:boost-density-complete-decomposition}
\end{align}
where 
\begin{equation}
\mathcal{H}^{i}{}_{k}:=-\frac{1}{2}x_{k}x^{i}\mathscr{R}+\frac{1}{4}\delta^{i}{}_{k}r^{2}\mathscr{R},\label{eq:boost-Hamiltonian-residual}
\end{equation}
is the weighted Hamiltonian-constraint contribution, and 
\begin{equation}
\mathcal{E}^{i}{}_{k}:=x_{k}x^{j}\mathscr{S}^{i}{}_{j}-\frac{1}{2}r^{2}\mathscr{S}^{i}{}_{k}\label{eq:boost-field-equation-residual}
\end{equation}
is the weighted spacetime field-equation contribution.

\subsubsection{The Regge--Teitelboim boost charge}

Because $\mathcal{Y}^{ij}{}_{k}$ is antisymmetric, 
\begin{equation}
\int_{S_{r}}\dd S_{i}\,\partial_{j}\mathcal{Y}^{ij}{}_{k}=0\label{eq:boost-Y-zero-flux}
\end{equation}
for every closed $S_{r}$, provided that the fields are regular in
a neighborhood of the asymptotic sphere. This is the integral of an
exact two-form over a closed surface and does not require any assumption
about the interior of the spacetime.

For asymptotically vacuum initial data, we require the weighted constraint
and field-equation remainders to satisfy 
\begin{equation}
\lim_{r\to\infty}\int_{S_{r}}\dd S_{i}\,\mathcal{H}^{i}{}_{k}=0\label{eq:boost-weighted-Hamiltonian-condition}
\end{equation}
and 
\begin{equation}
\lim_{r\to\infty}\int_{S_{r}}\dd S_{i}\,\mathcal{E}^{i}{}_{k}=0.\label{eq:boost-weighted-field-equation-condition}
\end{equation}
These conditions hold, in particular, when the asymptotic region is
vacuum to the order required by the boost charge. They are the weighted
versions of the asymptotic Hamiltonian constraint and the asymptotic
spacetime field equations. It follows from \eqref{eq:boost-density-complete-decomposition}
that 
\begin{align}
\lim_{r\to\infty}\int_{S_{r}}\dd S_{i}\,\mathcal{B}^{i}{}_{k}=\frac{1}{2}\lim_{r\to\infty}\int_{S_{r}}\dd S_{i}\Bigl[ & x_{k}\left(\partial_{j}h^{ij}-\partial^{i}h_{j}^{j}\right)-\left(h^{i}{}_{k}-\delta^{i}{}_{k}h_{j}^{j}\right)\Bigr].\label{eq:boost-flux-to-RT-density}
\end{align}
Substitution into \eqref{eq:boost-curvature-flux} yields 
\begin{align}
\mathcal{K}_{k}^{{\rm curv}}=\frac{1}{2\kappa_{4}}\lim_{r\to\infty}\int_{S_{r}}\dd S_{i}\Bigl[x_{k}\left(\partial_{j}h^{ij}-\partial^{i}h_{j}^{j}\right)-\left(h^{i}{}_{k}-\delta^{i}{}_{k}h_{j}^{j}\right)\Bigr].\label{eq:RT-boost-charge}
\end{align}
This is precisely the Regge--Teitelboim boost generator evaluated
on the $t=0$ hypersurface: 
\begin{equation}
\mathcal{K}_{k}^{{\rm curv}}=\mathcal{K}_{k}^{{\rm RT}}.\label{eq:curvature-RT-boost-equality}
\end{equation}

The subtraction term 
\begin{equation}
-\left(h^{i}{}_{k}-\delta^{i}{}_{k}h_{j}^{j}\right)\label{eq:RT-subtraction-term}
\end{equation}
has therefore not been inserted by hand. It is generated by the $-r^{2}(R^{i0}{}_{k0})^{(1)}/2$
contribution in the curvature density, together with the antisymmetric
improvements that arise when derivatives act on the boost weights.
This is precisely why the $-x^{2}\omega_{\beta\sigma}/2$ term in
the flat-space Killing potential is essential.

\subsubsection{Parity and finiteness}

The Regge--Teitelboim parity conditions are needed because the boost
Killing vector grows linearly with $r$. Write the spatial metric
perturbation as 
\begin{equation}
h_{ij}=\frac{a_{ij}(\hat{x})}{r}+\frac{b_{ij}(\hat{x})}{r^{2}}+o(r^{-2}),\qquad\hat{x}^{i}:=\frac{x^{i}}{r},\label{eq:boost-asymptotic-expansion}
\end{equation}
with 
\begin{equation}
a_{ij}(-\hat{x})=a_{ij}(\hat{x}).\label{eq:boost-leading-even-parity}
\end{equation}
The leading metric coefficient is therefore even under the antipodal
map. For the $a_{ij}/r$ term, $V^{i}$ is odd, whereas $x_{k}$ is
also odd. Consequently, 
\begin{equation}
x_{k}V^{i}-\left(h^{i}{}_{k}-\delta^{i}{}_{k}h_{j}^{j}\right),\label{eq:boost-leading-density-parity}
\end{equation}
is even. The directed area element 
\begin{equation}
\dd S_{i}=r^{2}\hat{x}_{i}\,\dd\Omega\label{eq:boost-directed-area-element}
\end{equation}
is odd. The nominally linearly divergent contribution is therefore
odd over the sphere and integrates to zero. The $r^{-2}$ coefficient
$b_{ij}$ supplies the finite boost charge. The leading ADM momentum
must simultaneously have odd parity, 
\begin{equation}
\Pi^{ij}=K^{ij}-K\gamma^{ij}=\frac{p^{ij}(\hat{x})}{r^{2}}+o(r^{-2}),\qquad p^{ij}(-\hat{x})=-p^{ij}(\hat{x}).\label{eq:boost-momentum-parity}
\end{equation}
Although the $t=0$ expression \eqref{eq:RT-boost-charge} is written
only in terms of the spatial metric, the momentum parity condition
is required for the complete canonical boost generator, its functional
differentiability, and the realization of the Poincaré algebra \cite{ReggeTeitelboim,BeigOMurchadha}.

\subsubsection{Displaced Schwarzschild check}

As a direct check, consider an asymptotically Schwarzschild field
whose center is displaced from the coordinate origin by a constant
vector $\boldsymbol{a}$. The spatial perturbation has the form 
\begin{equation}
h_{ij}=f\,\delta_{ij},\qquad f=\frac{A}{\lvert\boldsymbol{x}-\boldsymbol{a}\rvert}.\label{eq:translated-mass-asymptotics-start}
\end{equation}
At large $r$, 
\begin{equation}
f=\frac{A}{r}+\frac{A\,\boldsymbol{a}\cdot\hat{\boldsymbol{x}}}{r^{2}}+\order(r^{-3}).\label{eq:translated-mass-asymptotics}
\end{equation}
For Schwarzschild spacetime
\begin{equation}
A=2GM.\label{eq:translated-Schwarzschild-A}
\end{equation}
For this perturbation one can express the following results
\begin{equation}
h_{\mathrm{sp}}=3f,\qquad V^{i}=-2\partial^{i}f,\qquad h^{i}{}_{k}-\delta^{i}{}_{k}h_{\mathrm{sp}}=-2f\delta^{i}{}_{k}.\label{eq:translated-mass-basic-quantities}
\end{equation}
The ADM energy is 
\begin{equation}
E_{{\rm ADM}}=-\frac{1}{\kappa_{4}}\lim_{r\to\infty}\int_{S_{r}}\dd S_{i}\,\partial^{i}f=\frac{4\pi A}{\kappa_{4}}.\label{eq:translated-mass-energy}
\end{equation}
For $A=2GM$ and $\kappa_{4}=8\pi G$, this gives ADM energy, $E_{{\rm ADM}}=M$.
The boost charge becomes 
\begin{align}
\mathcal{K}_{k}=\frac{1}{\kappa_{4}}\lim_{r\to\infty}\int_{S_{r}}\dd S_{i}\left(-x_{k}\partial^{i}f+f\delta^{i}{}_{k}\right).\label{eq:translated-mass-boost-start}
\end{align}
Using 
\begin{equation}
\dd S_{i}=r^{2}\hat{x}_{i}\,\dd\Omega,\qquad x_{k}=r\hat{x}_{k},\label{eq:translated-mass-sphere-data}
\end{equation}
this becomes 
\begin{equation}
\mathcal{K}_{k}=\frac{1}{\kappa_{4}}\lim_{r\to\infty}\int_{S^{2}}\dd\Omega\,\hat{x}_{k}\left(-r^{3}\partial_{r}f+r^{2}f\right).\label{eq:translated-mass-boost-angular}
\end{equation}
From \eqref{eq:translated-mass-asymptotics} one obtains
\begin{equation}
-r^{3}\partial_{r}f+r^{2}f=2Ar+3A\,\boldsymbol{a}\cdot\hat{\boldsymbol{x}}+\order(r^{-1}).\label{eq:translated-mass-radial-combination}
\end{equation}
The first term of the right hand side last equation integrates to
zero because it is odd. Using 
\begin{equation}
\int_{S^{2}}\dd\Omega\,\hat{x}_{k}\hat{x}_{\ell}=\frac{4\pi}{3}\delta_{k\ell},\label{eq:unit-sphere-second-moment}
\end{equation}
the finite term gives 
\begin{equation}
\mathcal{K}_{k}=\frac{4\pi A}{\kappa_{4}}a_{k}=E_{{\rm ADM}}\ a_{k}.\label{eq:translated-mass-boost-result}
\end{equation}
Thus a mass centered at $\boldsymbol{a}$ has the expected boost charge.

\subsubsection{Boost charge versus center-of-mass position}

The conserved boost charge should be distinguished from the instantaneous
center-of-mass position. With the boost convention \eqref{eq:boost-xi},
the relation is 
\begin{equation}
\mathcal{K}_{k}=E_{{\rm ADM}}C_{k}(t)-tP_{k}.\label{eq:boost-charge-center-of-mass-relation}
\end{equation}
Therefore, when $E_{{\rm ADM}}\neq0$, 
\begin{equation}
C_{k}(t)=\frac{\mathcal{K}_{k}+tP_{k}}{E_{{\rm ADM}}}.\label{eq:center-of-mass-time-dependent}
\end{equation}
On the $t=0$ hypersurface used throughout the calculation, 
\begin{equation}
C_{k}(0)=\frac{\mathcal{K}_{k}}{E_{{\rm ADM}}}.\label{eq:center-of-mass}
\end{equation}
For the displaced Schwarzschild field, \eqref{eq:translated-mass-boost-result}
therefore gives 
\begin{equation}
C_{k}(0)=a_{k}.\label{eq:translated-mass-center-result}
\end{equation}
We have thus shown that the Regge--Teitelboim boost generator admits
a curvature-flux representative. The Regge--Teitelboim falloff and
parity conditions remain essential: they are not artifacts of the
curvature description, but the standard conditions required for finite,
coordinate-independent, and functionally differentiable Lorentz generators
at spatial infinity.

\subsection{Asymptotic Poincaré algebra}

\label{subsec:Poincare-algebra-RT}

With the Regge--Teitelboim boundary and parity conditions imposed,
the ten charges 
\begin{equation}
E,\qquad P_{i},\qquad J_{ij},\qquad\mathcal{K}_{i},\label{eq:Poincare-charge-list}
\end{equation}
are the differentiable canonical generators of the asymptotic Poincaré
group. The curvature-flux construction gives alternative representatives
of these same generators; it does not modify their canonical algebra.To
make the signs transparent, introduce the covariant notation 
\begin{equation}
P_{\mu}=(-E,P_{i}),\qquad J_{0i}=\mathcal{K}_{i},\qquad J_{ij}=-J_{ji}.\label{eq:Poincare-covariant-charge-identification}
\end{equation}
The boost convention $J_{0i}=\mathcal{K}_{i}$ agrees with 
\begin{equation}
\mathcal{K}_{i}=EC_{i}(t)-tP_{i},\label{eq:boost-generator-center-recalled}
\end{equation}
derived in the preceding subsection. We adopt the Hamiltonian-action
convention 
\begin{equation}
\delta_{\xi}F=\left\{ F,Q[\xi]\right\} .\label{eq:hamiltonian-action-convention}
\end{equation}
The Poincaré algebra may then be written covariantly as 
\begin{equation}
\{P_{\mu},P_{\nu}\}=0,\label{eq:Poincare-PP-covariant}
\end{equation}
\begin{equation}
\{J_{\mu\nu},P_{\rho}\}=\eta_{\mu\rho}P_{\nu}-\eta_{\nu\rho}P_{\mu},\label{eq:Poincare-JP-covariant}
\end{equation}
and 
\begin{align}
\{J_{\mu\nu},J_{\rho\sigma}\}={} & \eta_{\mu\rho}J_{\nu\sigma}-\eta_{\mu\sigma}J_{\nu\rho}-\eta_{\nu\rho}J_{\mu\sigma}+\eta_{\nu\sigma}J_{\mu\rho}.\label{eq:Poincare-JJ-covariant}
\end{align}

In terms of the spatial generators, the nonvanishing brackets are
\begin{align}
\{J_{ij},J_{kl}\}={} & \delta_{ik}J_{jl}-\delta_{il}J_{jk}-\delta_{jk}J_{il}+\delta_{jl}J_{ik},\label{eq:Poincare-JJ}\\
\{J_{ij},P_{k}\}={} & \delta_{ik}P_{j}-\delta_{jk}P_{i},\label{eq:Poincare-JP}\\
\{J_{ij},\mathcal{K}_{k}\}={} & \delta_{ik}\mathcal{K}_{j}-\delta_{jk}\mathcal{K}_{i},\label{eq:Poincare-JK}\\
\{\mathcal{K}_{i},E\}={} & P_{i},\label{eq:Poincare-KE}\\
\{\mathcal{K}_{i},P_{j}\}={} & \delta_{ij}E,\label{eq:Poincare-KP}\\
\{\mathcal{K}_{i},\mathcal{K}_{j}\}={} & -J_{ij}.\label{eq:Poincare-KK}
\end{align}
In addition, 
\begin{equation}
\{J_{ij},E\}=0,\qquad\{P_{i},E\}=0,\qquad\{P_{i},P_{j}\}=0.\label{eq:Poincare-remaining-zero-brackets}
\end{equation}
All remaining brackets follow from antisymmetry.

The standard normalization is chosen so that all ten generators vanish
on Minkowski initial data. This fixes their additive constants. Under
the Regge--Teitelboim falloff and parity conditions, the canonical
surface-term analysis then gives the algebra \eqref{eq:Poincare-PP-covariant}--\eqref{eq:Poincare-JJ-covariant}
without a central extension. The absence of a central term is therefore
a property of the differentiable canonical realization on the Regge--Teitelboim
phase space, rather than a consequence merely of the fact that the
individual charges vanish on Minkowski space.

\subsection{Example: Kerr spin as a curvature flux}

\label{subsec:Kerr-spin-curvature-flux}

We conclude the Lorentz-sector analysis with a direct normalization
check using the asymptotic Kerr field \cite{Kerr}. Let the angular
momentum point along the positive $z$-axis: 
\begin{equation}
\boldsymbol{J}=J\,\hat{\boldsymbol{z}},\qquad J_{12}=J_{z}=J.\label{eq:kerr-spin-z}
\end{equation}
We work to leading order in the spin-dependent asymptotic field. In
asymptotically Cartesian coordinates, the relevant gravitomagnetic
perturbation is 
\begin{equation}
h_{0i}=-2G\,\frac{\left(\boldsymbol{J}\times\boldsymbol{x}\right)_{i}}{r^{3}}.\label{eq:kerr-gravitomagnetic-general}
\end{equation}
For $\boldsymbol{J}=J\hat{\boldsymbol{z}}$, this gives 
\begin{equation}
h_{01}=\frac{2GJy}{r^{3}},\qquad h_{02}=-\frac{2GJx}{r^{3}},\qquad h_{03}=0.\label{eq:kerr-h0i-zspin}
\end{equation}

The stationary Schwarzschild part of the asymptotic metric has $h_{0i}=0$
and therefore does not contribute to the mixed curvature component
$R^{i0}{}_{kj}$ entering the rotational charge. At the order considered
here, the angular momentum is carried entirely by the spin-dependent
gravitomagnetic perturbation \eqref{eq:kerr-gravitomagnetic-general}.
For stationary perturbations, the linearized Riemann tensor gives
\begin{equation}
\left(R_{i0kj}\right)^{(1)}=\frac{1}{2}\partial_{i}\left(\partial_{j}h_{0k}-\partial_{k}h_{0j}\right).\label{eq:stationary-linearized-riemann-spin-lowered}
\end{equation}
Raising the first two indices with the Minkowski metric yields 
\begin{equation}
\left(R^{i0}{}_{kj}\right)^{(1)}=-\frac{1}{2}\partial^{i}\left(\partial_{j}h_{0k}-\partial_{k}h_{0j}\right).\label{eq:stationary-linearized-riemann-spin}
\end{equation}
For the rotation generator 
\begin{equation}
\xi_{(12)}=x\frac{\partial}{\partial y}-y\frac{\partial}{\partial x},\label{eq:kerr-z-rotation-vector}
\end{equation}
the curvature-flux expression \eqref{eq:angular-momentum-curvature-flux}
becomes 
\begin{align}
J_{12}^{{\rm curv}}=\frac{1}{\kappa_{4}}\lim_{r\to\infty}\int_{S_{r}}\dd S_{i}\Biggl[ & x\,x^{j}\left(R^{i0}{}_{2j}\right)^{(1)}-y\,x^{j}\left(R^{i0}{}_{1j}\right)^{(1)}+\frac{1}{2}r^{2}\left(R^{i0}{}_{12}\right)^{(1)}\Biggr].\label{eq:J12-curvature-flux}
\end{align}
Let 
\begin{equation}
x^{i}=r\hat{x}^{i},\qquad\hat{x}^{i}\hat{x}_{i}=1,\qquad\dd S_{i}=r^{2}\hat{x}_{i}\,\dd\Omega,\qquad u:=x^{3}=\cos\theta.\label{eq:kerr-sphere-data}
\end{equation}
Direct differentiation of \eqref{eq:kerr-h0i-zspin} gives the contribution
from the first two terms: 
\begin{align}
\dd S_{i}\Biggl[x\,x^{j}\left(R^{i0}{}_{2j}\right)^{(1)}-y\,x^{j}\left(R^{i0}{}_{1j}\right)^{(1)}\Biggr]=3GJ(1-u^{2})\,\dd\Omega.\label{eq:kerr-spin-main-integrand}
\end{align}
Its integral is 
\begin{align}
\int_{S^{2}}3GJ(1-u^{2})\,\dd\Omega & =6\pi GJ\int_{-1}^{1}(1-u^{2})\,\dd u=8\pi GJ.\label{eq:kerr-spin-main-integral}
\end{align}
The contribution of the Bianchi-potential term is locally nonzero:
\begin{equation}
\frac{1}{2}r^{2}\dd S_{i}\left(R^{i0}{}_{12}\right)^{(1)}=\frac{3GJ}{2}\left(3u^{2}-1\right)\dd\Omega.\label{eq:kerr-spin-correction-integrand}
\end{equation}
Its integrated flux vanishes because $3u^{2}-1$ is the $l=2$ trace-free
spherical harmonic: 
\begin{align*}
\frac{1}{2}\int_{S_{r}}\dd S_{i}\,r^{2}\left(R^{i0}{}_{12}\right)^{(1)} & =\frac{3GJ}{2}\int_{S^{2}}\left(3u^{2}-1\right)\dd\Omega=0.
\end{align*}
The fact that this term integrates to zero for the Kerr dipole does
not make it dispensable: it is required by the general Lorentz current
identity and is essential in the reduction to the canonical angular-momentum
expression.

Combining the two contributions, the complete curvature-flux density
may also be written as 
\begin{equation}
\dd S_{i}\,\mathcal{J}^{i}{}_{12}=\frac{3GJ}{2}\left(1+u^{2}\right)\dd\Omega.\label{eq:kerr-total-curvature-density}
\end{equation}
Therefore, 
\begin{align}
\lim_{r\to\infty}\int_{S_{r}}\dd S_{i}\,\mathcal{J}^{i}{}_{12} & =\frac{3GJ}{2}\int_{S^{2}}\left(1+u^{2}\right)\dd\Omega=8\pi GJ.\label{eq:kerr-spin-flux-integral-final}
\end{align}
Since $\kappa_{4}=8\pi G,$the curvature charge is 
\begin{equation}
J_{12}^{{\rm curv}}=J.\label{eq:J12-equals-J}
\end{equation}
For the Kerr black hole, 
\begin{equation}
J=Ma,\label{eq:kerr-J-Ma}
\end{equation}
and hence 
\begin{equation}
J_{z}^{{\rm curv}}=J_{z}^{{\rm ADM}}=Ma.\label{eq:kerr-curvature-ADM-spin-agree}
\end{equation}
The Kerr example therefore confirms both the normalization and the
sign of the rotational curvature-flux formula.

\section{Gauge invariance and boundary conditions}

\label{sec:gauge-invariance-boundary-conditions}

The curvature-flux representatives have an important advantage over
the usual metric-potential expressions: their dynamical parts are
manifestly invariant under proper linearized diffeomorphisms. This
local gauge invariance should, however, be distinguished from the
separate questions of finiteness, integrability, conservation, and
invariance of the integrated charge. Those questions depend on the
boundary conditions defining the asymptotic phase space.

\subsection{Linearized diffeomorphisms}

Let $X[g]$ be any tensor constructed covariantly from the metric
and its curvature. Covariance means that, for every diffeomorphism
$\phi$, one can express the following equality
\begin{equation}
X[\phi^{*}g]=\phi^{*}X[g].\label{eq:naturality-of-covariant-tensor}
\end{equation}
Consider the background split 
\begin{equation}
g_{\mu\nu}=\bar{g}_{\mu\nu}+h_{\mu\nu},\label{eq:gauge-section-background-split}
\end{equation}
with the background metric held fixed. An infinitesimal diffeomorphism
generated by $\zeta^{\mu}$ acts on the perturbation as 
\begin{equation}
\delta_{\zeta}h_{\mu\nu}=\mathcal{L}_{\zeta}\bar{g}_{\mu\nu}=2\bar{\nabla}_{(\mu}\zeta_{\nu)},\label{eq:linearized-gauge-transformation}
\end{equation}
where $\mathcal{L}_{\zeta}$ denotes the Lie derivative with respect
to vector field $\zeta$. Linearizing \eqref{eq:naturality-of-covariant-tensor}
gives the general identity 
\begin{equation}
\delta_{\zeta}X^{(1)}=\mathcal{L}_{\zeta}\bar{X}.\label{eq:linearized-tensor-gauge-transformation}
\end{equation}
Thus a linearized tensor is gauge invariant whenever the corresponding
background tensor vanishes or is invariant under arbitrary Lie differentiation
in the index placement being used.

\subsection{Flat background}

On the Minkowski background, 
\begin{equation}
\bar{R}_{\mu\nu\rho\sigma}=0.\label{eq:flat-background-Riemann-zero}
\end{equation}
Therefore, for the flat-space gauge transformation 
\begin{equation}
\delta_{\zeta}h_{\mu\nu}=\partial_{\mu}\zeta_{\nu}+\partial_{\nu}\zeta_{\mu},\label{eq:flat-small-gauge}
\end{equation}
equation \eqref{eq:linearized-tensor-gauge-transformation} gives
\begin{equation}
\delta_{\zeta}\left(R_{\mu\nu\rho\sigma}\right)^{(1)}=\mathcal{L}_{\zeta}\bar{R}_{\mu\nu\rho\sigma}=0.\label{eq:flat-Riemann-gauge-invariant}
\end{equation}
Because the background curvature vanishes, this statement holds for
every index placement of the linearized Riemann tensor. The same argument
applies to the flat-background $\mathcal{P}$-tensor. Namely, one
has
\begin{equation}
\bar{\mathcal{P}}^{\mu\nu}{}_{\rho\sigma}=0,\label{eq:flat-P-background-zero}
\end{equation}
and hence 
\begin{equation}
\delta_{\zeta}\left(\mathcal{P}^{\mu\nu}{}_{\rho\sigma}\right)^{(1)}=0.\label{eq:flat-P-linearized-gauge-invariant}
\end{equation}
The exact $\mathcal{P}$-tensor representative and its asymptotic
Riemann representative therefore both have gauge-invariant dynamical
integrands. The remaining quantities in the flat-space formulas, 
\begin{equation}
x^{\mu},\qquad\xi^{\mu}=a^{\mu}+\omega^{\mu}{}_{\nu}x^{\nu},\qquad F_{\mu\nu}[\xi],\label{eq:flat-fixed-background-weights}
\end{equation}
are fixed structures associated with the chosen asymptotic Cartesian
frame and the selected Poincaré generator. They are not varied when
the perturbation is subjected to a proper gauge transformation. Thus,
for an allowed gauge parameter $\zeta^{\mu}$ that decays sufficiently
rapidly at spatial infinity, 
\begin{equation}
\delta_{\zeta}Q[\xi]=0.\label{eq:flat-charge-proper-gauge-invariance}
\end{equation}
This statement does not imply invariance under a change of asymptotic
Poincaré frame. For example, a translation of the asymptotic origin
changes the numerical boost and angular-momentum components in the
standard way. The charges are well defined relative to the asymptotic
frame selected by the boundary conditions.

\subsection{Maximally symmetric AdS background}

On an AdS background, the gauge transformation of the linearized Riemann
tensor requires more care because the background curvature is nonzero.
With the mixed index placement used in the curvature charge, 
\begin{equation}
\bar{R}^{\nu\mu}{}_{\beta\sigma}=\frac{2\Lambda}{(n-1)(n-2)}\left(\delta^{\nu}{}_{\beta}\delta^{\mu}{}_{\sigma}-\delta^{\nu}{}_{\sigma}\delta^{\mu}{}_{\beta}\right).\label{eq:AdS-mixed-background-curvature}
\end{equation}
The right-hand side is a constant multiple of the identity tensor
on antisymmetric index pairs. Its Lie derivative therefore vanishes:
\begin{equation}
\mathcal{L}_{\zeta}\bar{R}^{\nu\mu}{}_{\beta\sigma}=0.\label{eq:AdS-mixed-background-Riemann-Lie-zero}
\end{equation}
Consequently, 
\begin{equation}
\delta_{\zeta}\left(R^{\nu\mu}{}_{\beta\sigma}\right)^{(1)}=\mathcal{L}_{\zeta}\bar{R}^{\nu\mu}{}_{\beta\sigma}=0.\label{eq:AdS-Riemann-gauge-invariant}
\end{equation}
This establishes the manifest gauge invariance of the dynamical curvature
appearing in \eqref{eq:AdS-Riemann-flux-charge-binormal}.

The mixed index placement is essential here. For example, the all-lowered
background curvature is 
\begin{equation}
\bar{R}_{\mu\nu\rho\sigma}=\frac{2\Lambda}{(n-1)(n-2)}\left(\bar{g}_{\mu\rho}\bar{g}_{\nu\sigma}-\bar{g}_{\mu\sigma}\bar{g}_{\nu\rho}\right),\label{eq:AdS-all-lowered-background-curvature}
\end{equation}
and therefore 
\begin{equation}
\delta_{\zeta}\left(R_{\mu\nu\rho\sigma}\right)^{(1)}=\mathcal{L}_{\zeta}\bar{R}_{\mu\nu\rho\sigma},\label{eq:AdS-all-lowered-Riemann-gauge-variation}
\end{equation}
does not vanish for a general $\zeta^{\mu}$. The variation contains
the Lie derivatives of the background metrics occurring in \eqref{eq:AdS-all-lowered-background-curvature}.
Thus it would be incorrect to claim that the linearized Riemann tensor
is gauge invariant in every index placement on AdS.

The shifted $\mathcal{P}$-tensor gives an even simpler argument.
On a maximally symmetric background, 
\begin{equation}
\bar{\mathcal{P}}^{\nu\mu}{}_{\beta\sigma}=0.\label{eq:Pbar-zero-gauge-section}
\end{equation}
It follows immediately that 
\begin{equation}
\delta_{\zeta}\left(\mathcal{P}^{\nu\mu}{}_{\beta\sigma}\right)^{(1)}=\mathcal{L}_{\zeta}\bar{\mathcal{P}}^{\nu\mu}{}_{\beta\sigma}=0.\label{eq:P-linear-gauge-invariant}
\end{equation}
Because the background value vanishes, this conclusion is independent
of the index placement. This is the cleanest proof of the gauge invariance
of the exact AdS $\mathcal{P}$-tensor representative \eqref{eq:AdS-P-charge}.

The background Killing vector and its derivative, 
\begin{equation}
\bar{\xi}^{\mu},\qquad\bar{\nabla}_{\mu}\bar{\xi}_{\nu},\label{eq:AdS-fixed-Killing-data}
\end{equation}
are held fixed in the linearized gauge transformation. Hence the complete
dynamical integrand in both the $\mathcal{P}$-tensor and mixed-Riemann
representatives is gauge invariant under proper transformations preserving
the asymptotic AdS phase space.

\subsection{Proper gauge transformations and asymptotic symmetries}

Not every diffeomorphism preserving the falloff conditions is a gauge
redundancy. One must distinguish two classes. A proper gauge transformation
is generated by a vector field $\zeta^{\mu}$ whose canonical surface
generator vanishes. Such a transformation changes the representative
$h_{\mu\nu}$ but not the physical point in the asymptotic phase space.
The charges are invariant: 
\begin{equation}
\delta_{\zeta}Q[\xi]=0.\label{eq:proper-gauge-charge-invariance}
\end{equation}

An asymptotic symmetry, by contrast, approaches a nontrivial Poincaré
or AdS Killing vector at infinity. Its surface generator is generally
nonzero. Such a transformation is not quotiented out as a gauge redundancy;
it acts nontrivially on the charges according to the corresponding
asymptotic symmetry algebra. Thus a Lorentz transformation may mix
the energy and momentum, and a translation of the origin may shift
the boost and angular momentum charges. These transformations do not
signal a failure of gauge invariance. Diffeomorphisms that violate
the chosen falloff or parity conditions do not preserve the phase
space under consideration. They are therefore neither proper gauge
transformations nor asymptotic symmetries of that phase space.

\subsection{Role of the boundary conditions}

Manifest gauge invariance of the local curvature does not by itself
ensure that a surface integral defines a gravitational charge. The
boundary conditions have several additional roles:
\begin{itemize}
\item First, they guarantee that the surface integral is finite. This is
particularly important for Lorentz generators, whose Killing vectors
grow linearly at spatial infinity. In four-dimensional asymptotically
flat spacetimes, the Regge--Teitelboim parity conditions eliminate
the nominally divergent terms in the rotation and boost integrals.
\item Second, the boundary conditions guarantee conservation: the charge
is independent of the hypersurface provided that the current has no
flux through the intervening asymptotic boundary.
\item Third, they determine whether the surface variation is integrable
in phase space and hence whether the charge defines a genuine Hamiltonian
generator.
\item Fourth, they justify the replacement of the exact $\mathcal{P}$-tensor
representative by a Riemann-flux representative. In flat space this
replacement requires the appropriately weighted Ricci and scalar-curvature
contributions to vanish at infinity. For Lorentz charges, the weights
grow quadratically, and the required decay is correspondingly stronger
than in the translation sector. In AdS, the analogous replacement
uses the standard asymptotically AdS falloff and the asymptotic field
equations.
\item Accordingly, the curvature-flux formulas do not replace the ADM, AD,
ADT, or Regge--Teitelboim definitions of the phase space and its
generators. Rather, after an admissible asymptotic phase space has
been specified, they provide curvature representatives of the same
conserved charges.
\item Finally, this discussion also clarifies why the flat-space formula
is not the $\Lambda\to0$ limit of the AdS formula. In AdS, the derivative
of a Killing vector is related algebraically to the Killing vector
itself through the nonzero constant background curvature. For a flat
translation, 
\begin{equation}
\partial_{\mu}a_{\nu}=0,\label{eq:flat-translation-Killing-form-zero-final}
\end{equation}
and that mechanism disappears. The flat construction must instead
use an antisymmetric Killing potential satisfying 
\begin{equation}
\partial_{\nu}F^{\nu\mu}[a]=a^{\mu}.\label{eq:flat-translation-potential-final}
\end{equation}
The AdS Killing two-form and the flat Killing potential therefore
play analogous roles, but they arise from genuinely different geometric
mechanisms.
\end{itemize}

\section{Generic Einstein backgrounds and the Weyl completion}

\label{sec:generic-Einstein-backgrounds}

We now ask to what extent the curvature-flux construction extends
from a maximally symmetric background to a generic Einstein background.
Two statements must be distinguished carefully.

First, the standard Abbott--Deser construction continues to provide
a codimension-two surface charge on an Einstein background admitting
a Killing vector. Second, the particularly simple and manifestly gauge-invariant
curvature representative obtained in AdS does not extend without additional
terms when the background Weyl tensor is nonzero. The obstruction
concerns the form of the representative, not the existence of the
conserved charge itself.

Let the background satisfy 
\begin{equation}
\bar{R}_{\mu\nu}=\frac{2\Lambda}{n-2}\,\bar{g}_{\mu\nu},\qquad\Lambda\neq0,\label{eq:Einstein-background-condition}
\end{equation}
but do not assume that it is maximally symmetric. Its scalar curvature
is 
\begin{equation}
\bar{R}=\frac{2n\Lambda}{n-2},\label{eq:Einstein-background-scalar}
\end{equation}
and its Riemann tensor decomposes as 
\begin{align}
\bar{R}_{\mu\nu\rho\sigma}=\bar{C}_{\mu\nu\rho\sigma}+\frac{2\Lambda}{(n-1)(n-2)}\left(\bar{g}_{\mu\rho}\bar{g}_{\nu\sigma}-\bar{g}_{\mu\sigma}\bar{g}_{\nu\rho}\right).\label{eq:Einstein-background-Riemann-decomposition}
\end{align}
For the shifted tensor introduced in \eqref{eq:P-cosmological-definition},
the background value is 
\begin{equation}
\bar{\mathcal{P}}^{\nu\mu}{}_{\beta\sigma}=\bar{C}^{\nu\mu}{}_{\beta\sigma},\qquad n>3.\label{eq:Pbar-equals-Weyl}
\end{equation}
Thus the shifted $\mathcal{P}$-tensor vanishes precisely when the
Einstein background is conformally flat and hence locally maximally
symmetric. Let $\bar{\xi}^{\mu}$ be a Killing vector, 
\begin{equation}
\bar{\nabla}_{(\mu}\bar{\xi}_{\nu)}=0,\label{eq:generic-Einstein-Killing-equation}
\end{equation}
and define its Killing two-form by 
\begin{equation}
\bar{S}^{\mu\nu}:=\bar{\nabla}^{\mu}\bar{\xi}^{\nu}=-\bar{\nabla}^{\nu}\bar{\xi}^{\mu}.\label{eq:Einstein-background-Killing-two-form}
\end{equation}

\subsection{The Killing potential still exists}

The existence of an antisymmetric Killing potential is not restricted
to a maximally symmetric background. The Killing identity and \eqref{eq:Einstein-background-condition}
give 
\begin{equation}
\bar{\nabla}_{\nu}\bar{S}^{\nu\mu}=\bar{\nabla}_{\nu}\bar{\nabla}^{\nu}\bar{\xi}^{\mu}=-\bar{R}^{\mu}{}_{\lambda}\bar{\xi}^{\lambda}=-\frac{2\Lambda}{n-2}\bar{\xi}^{\mu}.\label{eq:Einstein-Killing-form-divergence}
\end{equation}
Consequently, 
\begin{equation}
\bar{F}^{\nu\mu}[\bar{\xi}]:=-\frac{n-2}{2\Lambda}\,\bar{S}^{\nu\mu}\label{eq:generic-Einstein-Killing-potential}
\end{equation}
satisfies 
\begin{equation}
\bar{\nabla}_{\nu}\bar{F}^{\nu\mu}[\bar{\xi}]=\bar{\xi}^{\mu}.\label{eq:generic-Einstein-Killing-potential-divergence}
\end{equation}
The difference from AdS therefore does not arise from the absence
of a Killing potential. It arises from the nonzero background value
$\bar{\mathcal{P}}=\bar{C}$.

\subsection{Loss of background gauge invariance}

For any covariantly constructed tensor $X[g]$, a linearized diffeomorphism
generated by $\zeta^{\mu}$ acts according to 
\begin{equation}
\delta_{\zeta}X^{(1)}=\mathcal{L}_{\zeta}\bar{X}.\label{eq:generic-background-linearized-gauge-rule}
\end{equation}
It follows from \eqref{eq:Pbar-equals-Weyl} that 
\begin{equation}
\delta_{\zeta}\left(\mathcal{P}^{\nu\mu}{}_{\beta\sigma}\right)^{(1)}=\mathcal{L}_{\zeta}\bar{C}^{\nu\mu}{}_{\beta\sigma}.\label{eq:generic-Einstein-P-gauge-variation}
\end{equation}
This does not vanish for a generic gauge vector when $\bar{C}^{\nu\mu}{}_{\beta\sigma}\neq0$.
Thus the linearized shifted $\mathcal{P}$-tensor is not, by itself,
a background-gauge-invariant curvature perturbation on a generic Einstein
spacetime.

There is a second, closely related difference. Linearizing the exact
identity 
\begin{equation}
\nabla_{\nu}\mathcal{P}^{\nu\mu}{}_{\beta\sigma}=0,\label{eq:generic-Einstein-P-exact-divergence}
\end{equation}
about the Einstein background gives 
\begin{equation}
\bar{\nabla}_{\nu}\left(\mathcal{P}^{\nu\mu}{}_{\beta\sigma}\right)^{(1)}=-\mathcal{A}^{\mu}{}_{\beta\sigma}[h;\bar{C}],\label{eq:generic-Einstein-linearized-P-divergence}
\end{equation}
where 
\begin{align}
\mathcal{A}^{\mu}{}_{\beta\sigma}[h;\bar{C}]:={} & \left(\Gamma^{\nu}{}_{\nu\rho}\right)^{(1)}\bar{C}^{\rho\mu}{}_{\beta\sigma}-\left(\Gamma^{\rho}{}_{\nu\beta}\right)^{(1)}\bar{C}^{\nu\mu}{}_{\rho\sigma}-\left(\Gamma^{\rho}{}_{\nu\sigma}\right)^{(1)}\bar{C}^{\nu\mu}{}_{\beta\rho}.\label{eq:generic-Einstein-A-definition}
\end{align}
The term obtained by varying the second contravariant index vanishes
because $(\Gamma^{\mu}{}_{\nu\rho})^{(1)}$ is symmetric in $\nu,\rho$,
whereas $\bar{C}^{\nu\rho}{}_{\beta\sigma}$ is antisymmetric. The
tensor $\mathcal{A}^{\mu}{}_{\beta\sigma}$ is linear in $\bar{C}_{\mu\nu\rho\sigma}$
and in the first derivative of $h_{\mu\nu}$. It vanishes when the
background Weyl tensor vanishes.

Equation~\eqref{eq:generic-Einstein-linearized-P-divergence} shows
why the AdS calculation cannot simply be repeated: on a generic Einstein
background, the linearized $\mathcal{P}$-tensor is not divergence-free
with respect to the background derivative.

\subsection{The exact Weyl-corrected identity}

We can nevertheless determine exactly what replaces the AdS identity.
Consider 
\begin{equation}
\bar{\nabla}_{\nu}\left[\bar{S}^{\beta\sigma}\left(\mathcal{P}^{\nu\mu}{}_{\beta\sigma}\right)^{(1)}\right],\label{eq:generic-Einstein-product-start}
\end{equation}
using the product rule it gives 
\begin{align}
 & \bar{\nabla}_{\nu}\left[\bar{S}^{\beta\sigma}\left(\mathcal{P}^{\nu\mu}{}_{\beta\sigma}\right)^{(1)}\right]=\bar{S}^{\beta\sigma}\bar{\nabla}_{\nu}\left(\mathcal{P}^{\nu\mu}{}_{\beta\sigma}\right)^{(1)}+\left(\mathcal{P}^{\nu\mu}{}_{\beta\sigma}\right)^{(1)}\bar{\nabla}_{\nu}\bar{S}^{\beta\sigma}.\label{eq:generic-Einstein-product-rule}
\end{align}
The Killing identity is 
\begin{equation}
\bar{\nabla}_{\nu}\bar{S}^{\beta\sigma}=\bar{R}^{\sigma\beta}{}_{\nu\lambda}\bar{\xi}^{\lambda}.\label{eq:generic-Einstein-Killing-identity}
\end{equation}
Using the decomposition \eqref{eq:Einstein-background-Riemann-decomposition},
together with the trace identity $\left(\mathcal{P}^{\nu\mu}{}_{\nu\sigma}\right)^{(1)}=-(n-3)\left(\mathcal{G}^{\mu}{}_{\sigma}\right)^{(1)},$
one obtains 
\begin{align}
 & \bar{\nabla}_{\nu}\left[\bar{S}^{\beta\sigma}\left(\mathcal{P}^{\nu\mu}{}_{\beta\sigma}\right)^{(1)}\right]=\frac{4\Lambda(n-3)}{(n-1)(n-2)}\bar{\xi}_{\lambda}\left(\mathcal{G}^{\mu\lambda}\right)^{(1)}+\mathcal{W}^{\mu}[h;\bar{\xi}].\label{eq:generic-Einstein-exact-product-identity}
\end{align}
Here the complete background-Weyl current is 
\begin{align}
\mathcal{W}^{\mu}[h;\bar{\xi}]:={} & \left(\mathcal{P}^{\nu\mu}{}_{\beta\sigma}\right)^{(1)}\bar{C}^{\sigma\beta}{}_{\nu\lambda}\bar{\xi}^{\lambda}+\bar{S}^{\beta\sigma}\bar{\nabla}_{\nu}\left(\mathcal{P}^{\nu\mu}{}_{\beta\sigma}\right)^{(1)}.\label{eq:complete-Weyl-current}
\end{align}
Equivalently, using \eqref{eq:generic-Einstein-linearized-P-divergence},
\begin{align}
\mathcal{W}^{\mu}[h;\bar{\xi}]={} & \left(\mathcal{P}^{\nu\mu}{}_{\beta\sigma}\right)^{(1)}\bar{C}^{\sigma\beta}{}_{\nu\lambda}\bar{\xi}^{\lambda}-\bar{S}^{\beta\sigma}\mathcal{A}^{\mu}{}_{\beta\sigma}[h;\bar{C}].\label{eq:complete-Weyl-current-connection-form}
\end{align}
This expression contains both effects caused by the background Weyl
tensor: the Weyl part of the Killing identity and the failure of $(\mathcal{P}^{\nu\mu}{}_{\beta\sigma})^{(1)}$
to be background-divergence-free. Solving \eqref{eq:generic-Einstein-exact-product-identity}
for the Killing current gives 
\begin{align}
\bar{\xi}_{\lambda}\left(\mathcal{G}^{\mu\lambda}\right)^{(1)}={} & \frac{(n-1)(n-2)}{4\Lambda(n-3)}\Biggl\{\bar{\nabla}_{\nu}\left[\bar{S}^{\beta\sigma}\left(\mathcal{P}^{\nu\mu}{}_{\beta\sigma}\right)^{(1)}\right]-\mathcal{W}^{\mu}[h;\bar{\xi}]\Biggr\}.\label{eq:generic-Einstein-current-identity}
\end{align}
Equation~\eqref{eq:generic-Einstein-current-identity} is the exact
generalization of the AdS curvature identity. No unspecified remainder
has been omitted. When $\bar{C}_{\mu\nu\rho\sigma}=0$, one has 
\begin{equation}
\mathcal{W}^{\mu}[h;\bar{\xi}]=0,\label{eq:Weyl-current-zero-maxsym}
\end{equation}
and \eqref{eq:generic-Einstein-current-identity} reduces immediately
to the AdS result.

\subsection{The charge remains a surface charge}

The appearance of $\mathcal{W}^{\mu}$ does not imply that the conserved
charge ceases to possess a codimension-two representative. On every
Einstein background admitting a Killing vector, the standard Abbott--Deser
identity holds: 
\begin{equation}
\bar{\xi}_{\lambda}\left(\mathcal{G}^{\mu\lambda}\right)^{(1)}=\bar{\nabla}_{\nu}\mathcal{F}_{{\rm AD}}^{\mu\nu}[h;\bar{\xi}].\label{eq:generic-Einstein-AD-identity}
\end{equation}
Consequently, 
\begin{equation}
Q[\bar{\xi}]=\frac{1}{\kappa_{n}}\int_{\partial\bar{\Sigma}}\dd\bar{\Sigma}_{\mu\nu}\,\mathcal{F}_{{\rm AD}}^{\mu\nu}[h;\bar{\xi}]\label{eq:generic-Einstein-AD-charge}
\end{equation}
is a genuine codimension-two charge. Indeed, equations \eqref{eq:generic-Einstein-current-identity}
and \eqref{eq:generic-Einstein-AD-identity} show that the Weyl current
itself has a local superpotential. Define 
\begin{align}
\mathcal{V}_{W}^{\nu\mu}:={} & \bar{S}^{\beta\sigma}\left(\mathcal{P}^{\nu\mu}{}_{\beta\sigma}\right)^{(1)}+\frac{4\Lambda(n-3)}{(n-1)(n-2)}\mathcal{F}_{{\rm AD}}^{\nu\mu}[h;\bar{\xi}].\label{eq:Weyl-current-superpotential}
\end{align}
Then 
\begin{equation}
\mathcal{W}^{\mu}=\bar{\nabla}_{\nu}\mathcal{V}_{W}^{\nu\mu}.\label{eq:Weyl-current-is-divergence}
\end{equation}
Thus there is no obstruction to the existence of a surface charge.
The price is that the known superpotential $\mathcal{V}_{W}^{\nu\mu}$
contains the usual metric-potential term $\mathcal{F}_{{\rm AD}}^{\nu\mu}$.

The precise question is therefore not whether a generic Einstein charge
can be written as a surface flux. It can. The stronger question is
whether $\mathcal{V}_{W}^{\nu\mu}$ can be replaced by a universal
local expression constructed only from the linearized curvature, the
background curvature, the Killing vector, and finitely many of their
covariant derivatives.

No such universal curvature-only completion is established by the
present construction. The two immediate difficulties are 
\begin{equation}
\delta_{\zeta}\left(\mathcal{P}^{\nu\mu}{}_{\beta\sigma}\right)^{(1)}=\mathcal{L}_{\zeta}\bar{C}^{\nu\mu}{}_{\beta\sigma},\label{eq:generic-Einstein-curvature-gauge-obstruction}
\end{equation}
and 
\begin{equation}
\bar{\nabla}_{\nu}\left(\mathcal{P}^{\nu\mu}{}_{\beta\sigma}\right)^{(1)}=-\mathcal{A}^{\mu}{}_{\beta\sigma}[h;\bar{C}].\label{eq:generic-Einstein-curvature-divergence-obstruction}
\end{equation}
A curvature-only completion would have to cancel both effects while
remaining local, antisymmetric, gauge invariant, and compatible with
the required boundary conditions.

\subsection{Cases in which the curvature representative may survive}

Although the simple AdS expression is not valid automatically, it
may remain valid under additional assumptions. For example, suppose
that the background Weyl tensor and the perturbation obey boundary
conditions such that 
\begin{equation}
\lim_{\partial\bar{\Sigma}\to\infty}\int_{\partial\bar{\Sigma}}\dd\bar{\Sigma}_{\mu\nu}\,\mathcal{V}_{W}^{\nu\mu}=0.\label{eq:Weyl-completion-zero-boundary-flux}
\end{equation}
Then the Weyl completion makes no contribution at infinity, and the
charge is again represented by the AdS-type curvature flux 
\begin{align}
Q[\bar{\xi}]=\frac{(n-1)(n-2)}{4\kappa_{n}\Lambda(n-3)}\int_{\partial\bar{\Sigma}}\dd\bar{\Sigma}_{\mu\nu}\,\bar{S}^{\beta\sigma}\left(\mathcal{P}^{\nu\mu}{}_{\beta\sigma}\right)^{(1)}.\label{eq:generic-Einstein-asymptotic-curvature-flux}
\end{align}
Whether \eqref{eq:Weyl-completion-zero-boundary-flux} holds must
be checked for the particular background, perturbation, and asymptotic
phase space; it does not follow from the Einstein condition alone.

Additional simplifications may also occur for special Einstein backgrounds
possessing extra geometric structure, such as covariantly constant
curvature, Killing--Yano or conformal Killing--Yano tensors, Killing
spinors, or special algebraic type. Such structures can provide additional
curvature currents and superpotentials. They are background-specific,
however, and do not constitute a universal construction for arbitrary
Einstein spacetimes.

We therefore arrive at the following conclusion. Generic Einstein
backgrounds admitting Killing vectors possess the usual Abbott--Deser
codimension-two charges. The $\mathcal{P}$-tensor construction gives
the exact decomposition \eqref{eq:generic-Einstein-current-identity},
in which the departure from the AdS curvature flux is measured by
the explicit Weyl current \eqref{eq:complete-Weyl-current}. What
remains unresolved in full generality is not the existence of a surface
charge, but the existence of a universal, local, manifestly gauge-invariant,
curvature-only representative of its Weyl completion. This is an outstanding
problem which we have not been able to solve.

\section{Conclusions and outlook}

\label{sec:conclusions}

We have developed curvature-flux representatives of conserved gravitational
charges in two settings: perturbations about maximally symmetric AdS
backgrounds and asymptotically flat spacetimes. In both cases, the
construction is organized by the divergence-free rank-four tensor
$\mathcal{P}^{\nu\mu}{}_{\beta\sigma}$, whose algebraic symmetries
are those of the Riemann tensor and whose trace is proportional to
the cosmological Einstein tensor. The essential step is to contract
the linearized $\mathcal{P}$-tensor with an antisymmetric potential
adapted to the background Killing vector.

On a maximally symmetric AdS background, 
\begin{equation}
\bar{\mathcal{P}}^{\nu\mu}{}_{\beta\sigma}=0.\label{eq:conclusion-Pbar-zero-AdS}
\end{equation}
The linearized $\mathcal{P}$-tensor is consequently both background-divergence-free
and gauge invariant. Contracting it with the Killing two-form 
\begin{equation}
\bar{S}_{\beta\sigma}:=\bar{\nabla}_{\beta}\bar{\xi}_{\sigma},\label{eq:conclusion-AdS-Killing-two-form}
\end{equation}
converts the Abbott--Deser current into a total divergence. Under
the standard asymptotically AdS falloff conditions, the exact $\mathcal{P}$-tensor
representative reduces at infinity to 
\begin{equation}
Q[\bar{\xi}]=\frac{(n-1)(n-2)}{4\kappa_{n}\Lambda(n-3)}\int_{\partial\bar{\Sigma}}\dd\bar{\Sigma}_{\mu\nu}\,\left(R^{\nu\mu}{}_{\beta\sigma}\right)^{(1)}\bar{\nabla}^{\beta}\bar{\xi}^{\sigma},\qquad n>3.\label{eq:conclusion-AdS-Riemann-flux}
\end{equation}
This is not a new conserved quantity, but a manifestly curvature-based
representative of the standard Abbott--Deser charge. The above formula
can be extended to generic gravity theories via the methods used in
\cite{SenturkSismanTekin}.

The principal new result of the present work is the asymptotically
flat construction. It cannot be obtained by taking a naive $\Lambda\to0$
limit of the AdS formula. For a flat translation $a^{\mu}$, 
\begin{equation}
\partial_{\mu}a_{\nu}=0,\label{eq:conclusion-translation-gradient-zero}
\end{equation}
so the Killing two-form used in AdS vanishes. We instead introduced
an antisymmetric Killing potential satisfying 
\begin{equation}
\partial_{\nu}F^{\nu\mu}[\xi]=\xi^{\mu},\label{eq:conclusion-flat-Killing-potential-condition}
\end{equation}
for every Poincaré Killing vector 
\begin{equation}
\xi^{\mu}=a^{\mu}+\omega^{\mu}{}_{\nu}x^{\nu}.\label{eq:conclusion-Poincare-Killing-vector}
\end{equation}
The representative 
\begin{equation}
F^{\mu\nu}[\xi]=\frac{1}{n-1}\left(x^{\mu}\xi^{\nu}-x^{\nu}\xi^{\mu}\right)+\frac{1}{2(n-1)}x^{2}\omega^{\mu\nu},\label{eq:conclusion-flat-Poincare-potential}
\end{equation}
is adapted to the algebraic Bianchi identity. It yields the single
curvature-flux formula 
\begin{align}
Q[\xi]=\frac{1}{2\kappa_{n}(n-3)}\lim_{r\to\infty}\int_{S_{r}}\dd\Sigma_{\mu\nu}\,\left(R^{\nu\mu\beta\sigma}\right)^{(1)}\Bigl( & \xi_{\beta}x_{\sigma}-\xi_{\sigma}x_{\beta}-\frac{1}{2}x^{2}\omega_{\beta\sigma}\Bigr).\label{eq:conclusion-flat-master-charge}
\end{align}
The exact linearized identity is naturally written with the $\mathcal{P}$-tensor.
Equation~\eqref{eq:conclusion-flat-master-charge} is its asymptotic
Riemann representative, valid when the appropriately weighted Ricci
and scalar-curvature contributions vanish at spatial infinity. For
translations, \eqref{eq:conclusion-flat-master-charge} reproduces
the ADM four-momentum. In particular, 
\begin{align}
E & =\frac{1}{\kappa_{n}(n-3)}\lim_{r\to\infty}\int_{S_{r}}\dd S_{i}\,x^{j}\left(R^{i0}{}_{j0}\right)^{(1)},\label{eq:conclusion-flat-energy}\\
P_{k} & =\frac{1}{\kappa_{n}(n-3)}\lim_{r\to\infty}\int_{S_{r}}\dd S_{i}\,x^{j}\left(R^{i0}{}_{kj}\right)^{(1)}.\label{eq:conclusion-flat-momentum}
\end{align}
We showed directly that these expressions reduce to the standard ADM
surface integrals. Their normalization and signs were verified on
Schwarzschild and boosted Schwarzschild data, giving 
\begin{equation}
E=M,\qquad P_{i}=\gamma Mv_{i}.\label{eq:conclusion-translation-checks}
\end{equation}
In four dimensions, the same master formula gives the Lorentz charges.
The term proportional to $x^{2}\omega_{\beta\sigma}$ in \eqref{eq:conclusion-flat-Poincare-potential}
is essential: it is required for the Lorentz-dependent terms to collapse
by the algebraic Bianchi identity, and it generates the subtraction
term in the Regge--Teitelboim boost generator. With the usual Regge--Teitelboim
falloff and parity conditions, the curvature moments reproduce both
the ADM angular momentum and the boost charge. The Kerr check gives
\begin{equation}
J_{z}^{{\rm curv}}=J_{z}^{{\rm ADM}}=Ma,\label{eq:conclusion-Kerr-spin}
\end{equation}
while for a displaced Schwarzschild field one obtains 
\begin{equation}
\mathcal{K}_{k}=E_{{\rm ADM}}C_{k}(0).\label{eq:conclusion-displaced-Schwarzschild}
\end{equation}
More generally, 
\begin{equation}
\mathcal{K}_{k}=E_{{\rm ADM}}C_{k}(t)-tP_{k},\label{eq:conclusion-boost-center-relation}
\end{equation}
so the conserved boost generator should be distinguished from the
instantaneous center-of-mass position.

The curvature description does not eliminate the role of boundary
conditions. The boundary conditions define the phase space, distinguish
proper gauge transformations from asymptotic symmetries, guarantee
finiteness and integrability, and justify the replacement of the exact
$\mathcal{P}$-tensor expression by its asymptotic Riemann representative.
The curvature formulas should therefore be understood as alternative
representatives of the standard ADM, Abbott--Deser, and Regge--Teitelboim
charges within a fixed admissible phase space. On a generic Einstein
background, 
\begin{equation}
\bar{R}_{\mu\nu}=\frac{2\Lambda}{n-2}\bar{g}_{\mu\nu},\label{eq:conclusion-Einstein-background}
\end{equation}
the situation is different. For $n>3$, 
\begin{equation}
\bar{\mathcal{P}}^{\nu\mu}{}_{\beta\sigma}=\bar{C}^{\nu\mu}{}_{\beta\sigma},\label{eq:conclusion-Pbar-Weyl}
\end{equation}
and the bare linearized $\mathcal{P}$-tensor is therefore neither
pointwise gauge invariant nor, by itself, divergence-free with respect
to the background derivative. The unmodified AdS curvature representative
must consequently acquire Weyl-dependent completion terms.

This does not obstruct the existence of the charge. The usual Abbott--Deser
construction continues to provide a codimension-two metric-potential
representative on an Einstein background admitting a Killing vector.
What is not derived here is a universal, local, manifestly gauge-invariant
completion constructed only from the linearized curvature, the background
curvature, the Killing vector, and finitely many of their covariant
derivatives. Special Einstein backgrounds possessing additional geometric
structure may admit such curvature representatives, but no universal
construction for arbitrary Einstein backgrounds is claimed in this
work.

The central result may therefore be stated succinctly. On maximally
symmetric AdS backgrounds and on asymptotically flat backgrounds,
the standard conserved gravitational charges admit curvature-flux
representatives. In the flat case, the energy, momentum, angular momentum,
and boost or center-of-mass charges arise as different moments of
one linearized-curvature formula. Extending this unification to generic
Einstein backgrounds requires a controlled Weyl-dependent completion
and remains a natural problem for future work.

\appendix

\section{Canonical interpretation of the AdS curvature charge}

\label{app:canonical-generators-AdS}

In this appendix we clarify the canonical meaning of the AdS curvature-flux
formula. The curvature expression is an alternative representative
of the standard Abbott--Deser charge and therefore generates the
same asymptotic symmetry and obeys the same charge algebra.

We consider the standard global asymptotically AdS phase space in
$n>3$ dimensions, with fixed boundary conformal structure and boundary
conditions for which there is no symplectic flux through infinity.
The asymptotic symmetry algebra is then the finite-dimensional AdS
algebra 
\begin{equation}
\mathfrak{so}(n-1,2).\label{eq:appendix-AdS-algebra}
\end{equation}
For an asymptotic AdS Killing vector $\bar{\xi}^{\mu}$, decompose
\begin{equation}
\bar{\xi}^{\mu}=\bar{\xi}^{\perp}\bar{n}^{\mu}+\bar{\xi}^{i}e^{\mu}{}_{i},\label{eq:appendix-Killing-decomposition}
\end{equation}
where $\bar{n}^{\mu}$ is the future-directed normal to $\bar{\Sigma}$,
and $e^{\mu}{}_{i}$ is tangent to the hypersurface. The differentiable
canonical generator has the form 
\begin{equation}
G[\bar{\xi}]=\int_{\bar{\Sigma}}\dd^{\,n-1}x\left(\bar{\xi}^{\perp}\mathcal{H}+\bar{\xi}^{i}\mathcal{H}_{i}\right)+Q[\bar{\xi}],\label{eq:canonical-generator-AdS}
\end{equation}
where $\mathcal{H}$ and $\mathcal{H}_{i}$ are the Hamiltonian and
momentum constraints. The boundary term $Q[\bar{\xi}]$ is fixed by
requiring $G[\bar{\xi}]$ to possess well-defined functional derivatives.
On the constraint surface, 
\begin{equation}
\mathcal{H}=0,\qquad\mathcal{H}_{i}=0,\label{eq:constraints-vanish-AdS}
\end{equation}
and hence 
\begin{equation}
G[\bar{\xi}]\big|_{{\rm on\ shell}}=Q[\bar{\xi}].\label{eq:H-reduces-to-Q}
\end{equation}
For cosmological Einstein gravity, the boundary term may be represented
by the Abbott--Deser potential or, equivalently under the asymptotic
conditions used in this paper, by the curvature flux 
\begin{equation}
Q[\bar{\xi}]=c_{n}\int_{\partial\bar{\Sigma}}\dd\bar{\Sigma}_{\mu\nu}\,\left(R^{\nu\mu}{}_{\beta\sigma}\right)^{(1)}\bar{\nabla}^{\beta}\bar{\xi}^{\sigma},\label{eq:AdS-charge-algebra-start}
\end{equation}
where 
\begin{equation}
c_{n}:=\frac{(n-1)(n-2)}{4\kappa_{n}\Lambda(n-3)}.\label{eq:cn-definition}
\end{equation}
The Poisson algebra must be computed using the full generators \eqref{eq:canonical-generator-AdS}.
With the convention 
\begin{equation}
\delta_{\bar{\xi}}F=\{F,G[\bar{\xi}]\},\label{eq:Poisson-action-convention-AdS}
\end{equation}
the canonical algebra takes the form 
\begin{equation}
\{G[\bar{\xi}_{1}],G[\bar{\xi}_{2}]\}=G[[\bar{\xi}_{1},\bar{\xi}_{2}]]+K(\bar{\xi}_{1},\bar{\xi}_{2}),\label{eq:canonical-generator-algebra}
\end{equation}
where $K$ is a possible field-independent extension. Restricting
to the constraint surface gives 
\begin{equation}
\{Q[\bar{\xi}_{1}],Q[\bar{\xi}_{2}]\}=Q[[\bar{\xi}_{1},\bar{\xi}_{2}]]+K(\bar{\xi}_{1},\bar{\xi}_{2}).\label{eq:charge-algebra-with-central}
\end{equation}
For the standard normalization in which all charges vanish on the
AdS background, the extension can be evaluated on the background itself:
\begin{equation}
K(\bar{\xi}_{1},\bar{\xi}_{2})=\int_{\partial\bar{\Sigma}}k_{\bar{\xi}_{1}}\left[\mathcal{L}_{\bar{\xi}_{2}}\bar{g};\bar{g}\right],\label{eq:central-term-covariant-phase-space}
\end{equation}
where $k_{\bar{\xi}_{1}}$ is the standard covariant surface-charge
form. Because $\bar{\xi}_{2}^{\mu}$ is an exact Killing vector, 
\begin{equation}
\mathcal{L}_{\bar{\xi}_{2}}\bar{g}_{\mu\nu}=0,\label{eq:central-term-exact-Killing}
\end{equation}
and therefore 
\begin{equation}
K(\bar{\xi}_{1},\bar{\xi}_{2})=0.\label{eq:central-term-zero}
\end{equation}
Thus 
\begin{equation}
\{Q[\bar{\xi}_{1}],Q[\bar{\xi}_{2}]\}=Q[[\bar{\xi}_{1},\bar{\xi}_{2}]].\label{eq:AdS-charge-Poisson-algebra}
\end{equation}
This is the standard canonical realization of the AdS isometry algebra.
The absence of a central term is also consistent with the semisimplicity
of $\mathfrak{so}(n-1,2)$. The covariance of the curvature representative
provides an independent check of \eqref{eq:AdS-charge-Poisson-algebra}.
Since $\bar{\xi}_{2}^{\mu}$ is a background Killing vector, the perturbation
transforms as 
\begin{equation}
\delta_{\bar{\xi}_{2}}h_{\mu\nu}=\mathcal{L}_{\bar{\xi}_{2}}h_{\mu\nu}.\label{eq:h-transformation-background-Killing}
\end{equation}
The mixed-index linearized curvature consequently transforms tensorially:
\begin{equation}
\delta_{\bar{\xi}_{2}}\left(R^{\nu\mu}{}_{\beta\sigma}\right)^{(1)}=\mathcal{L}_{\bar{\xi}_{2}}\left(R^{\nu\mu}{}_{\beta\sigma}\right)^{(1)}.\label{eq:linearized-curvature-transformation}
\end{equation}
Moreover, if 
\begin{equation}
\bar{S}_{\beta\sigma}[\bar{\xi}]:=\bar{\nabla}_{\beta}\bar{\xi}_{\sigma},\label{eq:appendix-Killing-two-form}
\end{equation}
then 
\begin{equation}
\mathcal{L}_{\bar{\xi}_{2}}\bar{S}_{\beta\sigma}[\bar{\xi}_{1}]=\bar{S}_{\beta\sigma}\bigl[[\bar{\xi}_{2},\bar{\xi}_{1}]\bigr].\label{eq:Killing-two-form-Lie-transform}
\end{equation}
Together with the absence of flux through the AdS boundary, these
relations show directly that the curvature charge transforms in the
adjoint representation:
\begin{equation}
\delta_{\bar{\xi}_{2}}Q[\bar{\xi}_{1}]=Q[[\bar{\xi}_{1},\bar{\xi}_{2}]].\label{eq:curvature-charge-adjoint-transformation}
\end{equation}
To display the algebra explicitly, let $\bar{\xi}_{AB}=-\bar{\xi}_{BA}$
be a basis of AdS Killing vectors, with $A,B=0,1,\ldots,n,$ and define
\begin{equation}
J_{AB}:=Q[\bar{\xi}_{AB}].\label{eq:AdS-ambient-charges}
\end{equation}
Then 
\begin{align}
\{J_{AB},J_{CD}\}={} & \eta_{AC}J_{BD}-\eta_{AD}J_{BC}-\eta_{BC}J_{AD}+\eta_{BD}J_{AC}.\label{eq:so-nminus1-2-charge-algebra}
\end{align}
The number of independent generators is $\frac{n(n+1)}{2}.$ Relative
to a chosen global AdS frame, these generators decompose into the
energy, angular momenta, and the remaining AdS boost or transvection
charges.

This finite-dimensional result should be distinguished from the Brown--Henneaux
\cite{Brown} algebra in three-dimensional gravity. Under Brown--Henneaux
boundary conditions, the exact $\mathfrak{so}(2,2)$ isometry algebra
is enlarged to two copies of the Virasoro algebra generated by asymptotic,
rather than exact, Killing vectors. For such generators, 
\begin{equation}
\mathcal{L}_{\xi}\bar{g}_{\mu\nu}\neq0,\label{eq:asymptotic-not-exact-Killing}
\end{equation}
and the background evaluation 
\begin{equation}
\int_{\partial\bar{\Sigma}}k_{\xi_{1}}\left[\mathcal{L}_{\xi_{2}}\bar{g};\bar{g}\right]\label{eq:Brown-Henneaux-central-computation}
\end{equation}
need not vanish. This is the origin of the Brown--Henneaux central
charge.

For the standard $n>3$ global asymptotically AdS phase space considered
here, the boundary conformal group is finite-dimensional and coincides
with the AdS group. The curvature-flux construction does not enlarge
this asymptotic symmetry algebra. Its role is instead to provide one
gauge-invariant curvature representative for all the standard AdS
charges.

\section*{Acknowledgments}

We thank Hamed Adami for useful discussions. Details of the computations
are available at Zenodo \cite{Altas-Zenodo}. We dedicate this work to the memory
of Sergey Borisenok, who tragically passed away on 20 July 2026.

\end{document}